%% file: journalpaper.tex
\documentclass[preprint,authoryear,5p,times,twocolumn]{elsarticle}
\usepackage{lineno}
\modulolinenumbers[5]
\usepackage{hyperref}
\usepackage{array,multirow,graphicx}
\usepackage{amsmath}
\usepackage{xcolor}

\usepackage{calc}
\usepackage{enumitem}
\usepackage{tabularx,booktabs,multirow,makecell} 
\usepackage{ifpdf}
\ifpdf
 \usepackage{graphicx}
 \DeclareGraphicsExtensions{.pdf}
\else
 \usepackage[dvips]{graphicx}
 \DeclareGraphicsExtensions{.eps}
\fi
\graphicspath{{./images/}}

\usepackage{tikz}
\usetikzlibrary{shapes, shadows, arrows}
\usetikzlibrary{positioning}

\journal{Astronomy and Computing}

\newcommand{\kb}[1]{{\color{blue}{#1}}} 

\renewcommand{\vec}[1]{\boldsymbol{#1}}

\begin{document}

\begin{frontmatter}

\title{Detection of extragalactic Ultra-Compact Dwarfs and Globular Clusters using Explainable AI techniques}

\author[mysecondaryaddress]{\texorpdfstring{Mohammad Mohammadi\corref{mycorrespondingauthor}\fnref{myfootnote}}}
\ead{mohammadimathstar@gmail.com}
\author[mysecondaryaddress]{\texorpdfstring{Jarvin Mutatiina\fnref{myfootnote}}}
\author[kapteyn]{\texorpdfstring{Teymoor Saifollahi\fnref{myfootnote}}}
\author[mysecondaryaddress]{Kerstin Bunte}
\ead{kerstin.bunte@googlemail.com}
\ead[url]{www.cs.rug.nl/\string~kbunte}
\address{Faculty of Science and Engineering, University of Groningen, The Netherlands}
\cortext[mycorrespondingauthor]{Corresponding author}
\fntext[myfootnote]{These authors contributed equally.}

\address[mysecondaryaddress]{Bernoulli Institute for Mathematics, Computer Science and Artificial Intelligence}
\address[kapteyn]{Kapteyn Astronomical Institute, University of Groningen, the Netherlands}


%
%

\begin{abstract}

Compact stellar systems such as Ultra-compact dwarfs (UCDs) and Globular Clusters (GCs) around galaxies are known to be the tracers of the merger events that have been forming these galaxies. Therefore, identifying such systems allows 
to study galaxies mass assembly, formation and evolution. 
However, in the lack of spectroscopic information detecting UCDs/GCs using imaging data is very uncertain. 
Here, we aim to train a machine learning model 
to separate these objects from the foreground stars and background galaxies using the multi-wavelength 
imaging data of the Fornax galaxy cluster in 6 filters, namely \emph{u, g, r,  i, J} and \emph{Ks}. The classes of objects are highly imbalanced which is problematic for many automatic classification techniques. Hence, we employ Synthetic Minority Over-sampling 
to handle the imbalance of the training data. Then, we compare two classifiers, namely Localized Generalized Matrix Learning Vector Quantization (LGMLVQ) and Random Forest (RF). Both methods are able to identify UCDs/GCs with a precision and a recall 
of $>93\%$ and provide relevances that reflect the importance of each feature dimension 
for the classification. Both methods detect angular sizes as important markers for this classification problem. While it is astronomical expectation that color indices of $u-i$ and $i-Ks$ are the most important colors, our analysis shows that colors such as $g-r$ are more informative, potentially because of higher signal-to-noise ratio. Besides the excellent performance the LGMLVQ method allows further interpretability by providing the feature importance for each individual class, class-wise representative samples and the possibility for non-linear visualization of the data as demonstrated in this contribution. We conclude that employing machine learning techniques to identify UCDs/GCs can lead to promising results. Especially transparent methods 
allow further investigation and analysis of importance of the measurements for the detection problem and provide tools for 
non-linear visualization of the data.
\end{abstract}


\end{frontmatter}


\input{sections/introduction.tex}
\input{sections/data.tex}
\input{sections/methods.tex}
\input{sections/experiments.tex}
\input{sections/conclusion.tex}

\subsection*{Acknowledgments}
This project has received financial support from the European Union's Horizon 2020 research and innovation program under the Marie Sk{\l}odowska-Curie grant agreement No. 721463 to the SUNDIAL ITN network.

\bibliographystyle{model2-names}
\bibliography{bibfile}

\end{document}

%% file: sections/introduction.tex
\section{Introduction}
\label{sec:introduction}
Based on the current theories of galaxy formation and evolution \citep*{galaxyformation}, galaxies are formed hierarchically from the merger of low-mass galaxies that were formed earlier.
In this picture, the dense stellar structures such as Ultra-Compact Dwarf (UCD) galaxies and Globular Clusters (GCs), which are mostly found around galaxies and the core of galaxy clusters, are known to be the tracers of such merging events \citep{beasley2020}. However, extragalactic UCDs and GCs around other galaxies than the Milky Way look like stars (point-sources), due to their distance and current limitations of observational equipment. Therefore identifying them through imaging is challenging \citep{jordan2009}. 
To find these objects, it is necessary to measure their distances, which is only possible using spectroscopy and measuring their radial velocity. 
The Hubble-Lema\^{\i}tre Law \citep{hubble,lemaitre} says that more distant galaxies move faster away from us and thus astronomers measure the distance of galaxies by measuring their radial velocity. The latter can be measured using Doppler shifts of the absorption lines in the spectroscopic observations. Spectroscopy of astronomical objects, however, needs longer exposure times than imaging.  In other words, spectroscopy for all the star-like objects (point-sources) is unfortunately not feasible in practice \citep{voggel2020}. 

The recent advances in astronomical instrumentation and observations, without doubt, have provided us with a large amount of data to explore. 
Traditionally, the possible UCDs/GCs candidates are identified by multi-wavelength observations in a few optical filters \citep{Cantiello2018}. 
Once the candidates are found follow-up spectroscopy for selected nominees is carried out to measure the radial velocity and therefore the distance to confirm the 
identity of the objects \citep{pota}. This implies that a more accurate UCD/GC selection makes the observation time shorter. 
Recently, \citealp{munoz} has shown that a combination of optical/near infrared filters improves the quality of identifying UCDs/GCs. 
However, this approach was not used until very recently, mostly because deep observations in the near-infrared are not as easy as the optical.
In the case of Fornax galaxy cluster, the second closest galaxy cluster to us, recent observations of optical and near-infrared have been available, which makes it possible to identify UCDs/GCs within the galaxy cluster. 
The optical part of the data has been used earlier to identify GCs in the cluster using various techniques such as Bayesian Mixture Models, Growing Neural
Gas model and K-nearest neighbours \citep{DAbrusco,Prole-2019,angora,Cantiello2020,teymoor}. 
%

Due to the sheer amount of astronomical data, automated 
tools for analysis 
are highly desirable.
Therefore, machine learning techniques get more and more attention among astronomers recently, and they have been popularly explored for astronomical applications.
The Random Forest (RF) has been used to build a classifier for quasar identification \citep{carrasco2015photometric, gao2009random} and the Support Vector Machine (SVM) has been employed to estimate the redshift \citep{Jones_2017}.
Other techniques used are k-nearest neighbor classifier for active galactic nuclei (AGN) detection \citep{li2008k},
Support Vector Data Description (SVDD) for GC detection \citep{mohammadi2019globular}, Multi Layer Perceptron (MLP) for estimating star formation rate \citep{delli2019star}, Linear Discriminant Analysis (LDA) for identification of galaxy mergers \citep{nevin2019accurate}, and Artificial Neural Networks (ANN) for galaxy morphological classification and AGN detection \citep{ball2004galaxy,barchi2020machine,xiao2020efficient}.
Thus, a pletora of machine learning methods have been successfully applied in varying astronomical applications signifying a growing interest and mutually beneficial synergy. 

Explainable Artificial Intelligence (XAI) generally includes techniques which provide output that
can be interpreted by a human. 
Two examples of XAI techniques for classification are Learning Vector Quantization (LVQ) and Random Forest (RF) \citep{breiman2001random}. 
The latter bags an ensemble of decision trees (DTs) for classification. 
The trees within, are constructed from bootstrap examples that are sampled independently with replacement and random feature selection, 
following a common distribution while growing to maximum depth with no pruning until all leaves are pure. 
The label for a given input is predicted according to the most popular predicted label among all trees.
The stochastic ensemble strategy is robust to outliers and improves the performance, because of the law of large numbers from combination of rather low performing/weak DT classifiers \citep{breiman2001random}. 
LVQ comprises a well-known family of prototype-based classifiers 
that can efficiently distinguish potentially high dimensional multi-class problems. 
One of the main advantages of LVQ classifiers is the interpretability of the adaptive parameters. 
The learned prototypes, for example, can be investigated as typical representatives of the classes. 
While the original formulations of the LVQ family, such as \emph{Generalized learning Vector Quantization} (GLVQ) \citep{sato&yamada} use the Euclidean distance, more complex extensions, such as Generalized Matrix LVQ (GMLVQ) and the localized version LGMLVQ \citep{schneiderBiehl, petrabiehl} employ adaptive distance metrics. 
The latter algorithms also allow further insights by visualization of the decision boundaries after training \citep{BUNTELiram, kbuntethesis}. 

In this contribution, we use an ensemble of LGMLVQ and RF to classify three astronomical structures, namely foreground stars, UCDs/GCs and background galaxies, based on their optical ($u$, $g$, $r$, $i$) and near-infrared ($J$, $Ks$) measurements of the Fornax Deep Survey, VISTA Hemisphere Survey and ESO/VISTA archive. 
The LVQ and RF methods construct non-linear decision boundaries and the former allows to evaluate the importance of features for each class individually. 
One major issue often faced in astronomical problems is the imbalance of the classes. 
The total number of known UCDs and GCs in the Fornax cluster identified in the data is just over 500, whereas the majority class contains about 5 times more instances. 
To tackle this challenge we apply over-sampling techniques, such as Synthetic Minority Oversampling (SMOTE) \citep{smotechawla} and Borderline SMOTE \citep{borderlinesmote}. 
In contrast to previous works we have both optical and near-infrared filters in the dataset. 
We use an ensemble of LGMLVQ and RF to detect the classes of objects in large amounts of high dimensional astronomical data, compare the performances and analyse the results, by detailing important features and visualization. 

The paper is organized as follows: 
In section \ref{sec:data} we describe the dataset followed by the explanation of 
classifiers in section \ref{sec:methods}. 
Afterwards, in section \ref{sec:experiments} we describe the experiments and discuss the results. 
Finally we conclude in section \ref{sec:conclusion} and provide inspirations for future work.

%% file: sections/data.tex
\section{Astronomical Data and Preprocessing}
\label{sec:data}

The data used in this research is extracted from multi-wavelength wide astronomical surveys obtained from a combination of 6 filters, 
i.e optical filters ($u$, $g$ ,$r$ and $i$) and near-infrared filters ($J$ and $Ks$). 
The optical $u$, $g$, $r$, $i$ data was obtained from Fornax Deep Survey (FDS) using the ESO VLT survey telescope (VST), 
$J$ from VISTA Hemisphere Survey (VHS, \citealp{vhs}) using the VISTA telescope and $Ks$ from the ESO/VISTA archival data. 
The imaging data in $u$, $g$, $r$, $i$, $J$, $K_s$ has 5$\sigma$ limiting magnitude (point-source detection with signal to noise ratio S/N = 5) about 24.1, 25.4, 24.9 and 24.0, 20.7, 18.4 mag respectively. 
The limiting magnitudes in optical ($ugri$) and near-infrared ($JK_s$) are expressed in AB and Vega magnitude systems.
The data set provides photometric information in the direction of the Fornax galaxy cluster and is described in detail in \citet{teymoor}.

Here we use the catalogue of the spectroscopically confirmed objects referred to as $KNOWN$ catalogue in \citet{teymoor},
that consists of sources with existing radial velocity value in the literature.
After excluding larger objects likely to be galaxies in the Fornax cluster or slightly more distant and removing UCDs/GCs that are closer than 60 arcsec to the brightest Fornax cluster galaxies, mainly NGC1399 (to avoid larger uncertainties in colours and magnitudes),
any observed object in the dataset belong to one of the following 3 classes:
\begin{enumerate}[topsep=0pt,noitemsep,label=class \arabic*:,partopsep=0pt,parsep=0pt,labelwidth=\widthof{\ref{class3}},leftmargin=!]
\item 
consists of 2826 background galaxies further away than the galaxies in the Fornax cluster,
\item 
denotes the class of interest consisting of 512 UCDs and GCs, and 
\item 
contains 4399 nearby foreground stars located in our own galaxy, the Milky Way.
\label{class3}
\end{enumerate}
In the absence of spectroscopic data, these classes are difficult to distinguish for two main reasons: 
(1) the UCD and GC (class 2) observations are faint and ambiguous
as they are engulfed between the two other classes, and (2) there is only a small number of confirmed examples of them. 

\begin{figure*}
\centering
\includegraphics[width=\textwidth]{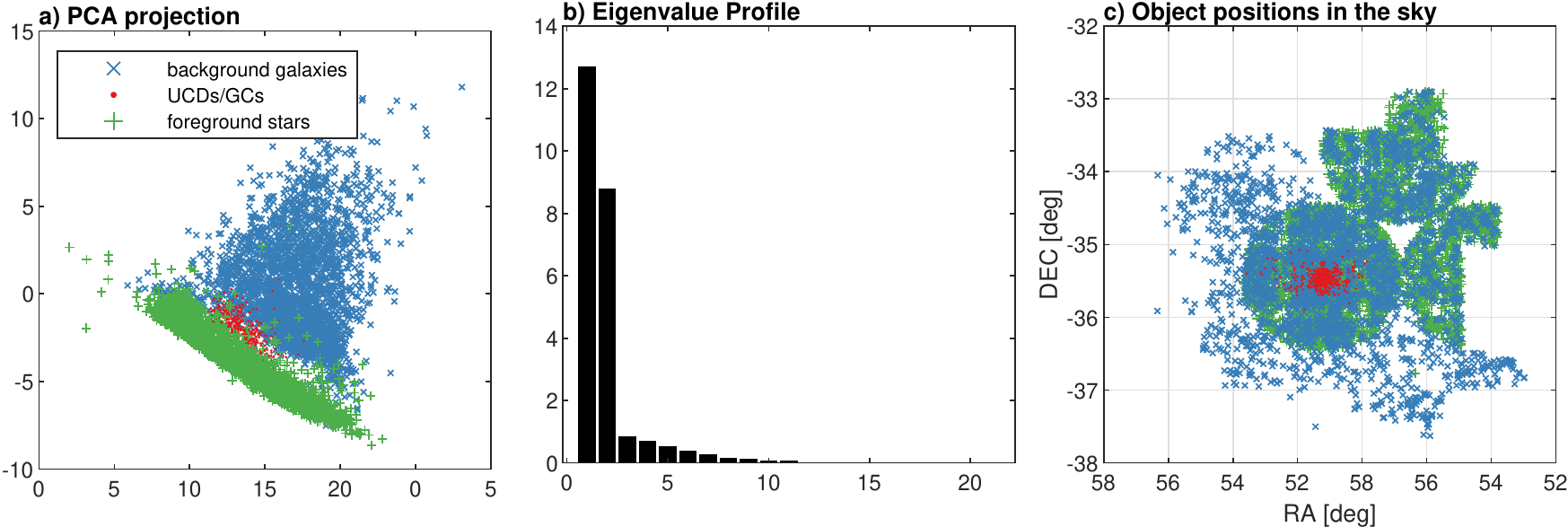}
\caption{
The position of the objects in the sky in right ascension (RA) and declination (DEC) (panel c).
PCA-projection of the data colouring the background galaxies, UCDs/GCs and foreground stars, 
and its corresponding Eigenvalue profile (panel a and b), respectively. 
}
\label{fig: datadefinition}
\end{figure*}

The labelled data consists of 23 features. 
The coordinates of the objects in the sky are given in right ascension ($RA$) and declination ($DEC$) as a degree of point of observation from earth and illustrated in Fig.~\ref{fig: datadefinition}(c). 
Features $FWHM^{\ast}g$, $FWHM^{\ast}r$, $FWHM^{\ast}i$, $FWHM^{\ast}u$, $FWHM^{\ast}j$ and $FWHM^{\ast}k$, also known as the Full width half maximum,
are the proxies for the angular sizes of the objects as seen from the corresponding filters. 
Followed by $u-g$, $u-r$, $u-i$, $u-J$, $u-Ks$, $g-r$, $g-i$, $g-J$, $g-Ks$, $r-i$, $r-J$, $r-Ks$, $i-J$, $i-Ks$ and $J-Ks$, which are 
colour indices indicating the emission of the astronomical object in logarithmic scale, known as magnitude and correlated to physical properties like age and metallicity. 
The FWHM$^{\ast}$ parameter constitutes an alternative way to estimate the compactness of sources and consequently it is similar to the other measures, such as the concentration index.
The data used in our analysis consists of 7737 complete observations in total.

Figure~\ref{fig: datadefinition}(b) shows the Eigenvalue profile of the data obtained using the unsupervised Principal Component Analysis (PCA). 
The two major Eigenvalues explain 88\% of the data's variance and the corresponding two-dimensional projection is depicted in panel (a).
The large areas of overlap and imbalance of the classes is clearly visible. 
Especially the latter states a problem for most classification techniques and 
thus re-sampling techniques, such as Synthetic Minority Over-sampling (SMOTE) \citep{smotechawla} and Borderline-SMOTE \citep{borderlinesmote}, 
are investigated as preprocessing steps. 
SMOTE increases the population of the minority classes by generating synthetic samples as weighted convex combination between random samples and its nearest neighbours. 
Instead of choosing random samples Borderline-SMOTE specifically takes points near the boundaries between the classes which are more likely to be misclassified. 
In this contribution we investigate and compare both methods as preprocessing to balance the classes. 

%% file: sections/methods.tex
\section{Methods}
\label{sec:methods}
In this section, we introduce the state-of-the-art methodology used, with focus on localized adaptive distance metrics in Learning Vector Quantization (LVQ), 
and corresponding interpretability. 
Moreover, we present a comparison between Random Forest (RF) (which is an ensemble of Decision Trees) and an ensemble of the Learning Vector Quantization models.

\subsection{Learning Vector Quantization (LVQ)}

We assume $\{(\vec{\xi}_i,y_i)\}_{i=1}^{n} $ denote the training set, where  $\vec{\xi}_i\in \mathbb{R}^N$ and 
$y_i \in \{1,\cdots,C\}$ represent $i$-th data point and its class label, respectively. 
An LVQ classifier models the distribution of classes via a set of labelled prototypes $\{ (\vec{\omega^j}, c(\vec{\omega^j})) \}_{j=1}^m$, 
where $c(\vec{\omega^j})$ is the label of the respective prototype. 
These prototypes tessellate the data space into smaller regions, called Voronoi cells, enclosing data points for which the respective prototype is closer than any other.
Classification follows a \emph{nearest prototype scheme}, meaning any data point (including new ones) is assigned the class label of the nearest prototype. 
To find prototype positions that minimize the classification error $E$, \emph{Generalized learning Vector Quantization} (GLVQ) \citep{sato&yamada} 
introduced the following cost function, aiming at large margin optimization for better generalization:
\begin{equation}
\label{eq:glvq}
E = \sum_{i=1}^n \Phi(\mu_i) \quad \mathrm{with } \quad \mu_i = \frac{d(\vec{\xi}_i, \vec{\omega}^J) - d(\vec{\xi}_i, \vec{\omega}^K)}
{d(\vec{\xi}_i, \vec{\omega}^J) +d(\vec{\xi}_i, \vec{\omega}^K)} \enspace,
\end{equation}
with $\Phi$ being a monotonically increasing function. We used the identity $\Phi(x) = x$ throughout this contribution.
Furthermore, $d(\vec{\xi}_i, \vec{\omega}^J)$ denotes the distance of the data point $\vec{\xi}_i$ from the closest prototype with the same label $y_i = c(\vec{\omega}^J)$ and 
$d(\vec{\xi}_i, \vec{\omega}^K)$ the distance to the closest prototype with a different class label $y_i\neq c(\vec{\omega}^K)$. 
The value of $\mu_i \in [-1,1]$ can be understood as a measure of confidence for the classification of sample $\vec{\xi}_i$. 
The closer $\mu_i$ is to -1, the smaller the distance 
to the closest prototype with the same label compared to the distance to the closest prototype with a different label, i.e.
$d(\vec{\xi}_i, \vec{\omega}^J)\ll d(\vec{\xi}_i, \vec{\omega}^K)$, 
the more certain the classification 
of $\vec{\xi}_i$.
More formally, one can compute the probability of a sample $\vec{\xi}_i$ to belong to class $j$, based on the distances to the prototypes, using the softmax:
\begin{equation}
\label{eq:softmax}
P(j|\vec{\xi}_i) = \frac{\exp{(-d(\vec{\xi}_i, \vec{\omega}^j))}}{\sum_{k=1}^{C} \exp{(-d(\vec{\xi}_i, \vec{\omega}^k))}} \enspace .
\end{equation}
The cost function is non-convex and typically gradient techniques, such as stochastic gradient descent \citep{petrabiehl,kbuntethesis}, are utilized to minimize the costs 
Eq.~\kb{\eqref{eq:glvq}}.

From the cost function Eq.\ \eqref{eq:glvq} it is clear that the the distance measure $d$ plays a major role for the performance of LVQ classifiers. 
While the Euclidean distance is a common choice all dimensions contribute equally in it, which has 
drawbacks in capturing underlying data semantics in noisy high-dimensional and heterogeneous data spaces \citep{petrabiehl}. 
As such it is not capable to reflect if features differ in importance for the classification task at hand.
Therefore, \citet{hammer2002generalized} proposed to incorporate a weighting factor for each feature dimension that is adapted during training: 
\begin{equation}
\label{eq:adaptmetric}
d^{\Lambda} (\vec{\xi}_i, \vec{\omega}^j) = (\vec{\xi}_i-\vec{\omega}^j)^\top \Lambda (\vec{\xi}_i-\vec{\omega}^j) \enspace,
\end{equation}
where the weight matrix $\Lambda$, also referred to as \emph{relevance matrix}, is a diagonal matrix with $0$ in the off-diagonals and 
$\lambda_i$ on the diagonal with $\sum_i \lambda_i=1$. 
These relevance weights indicate the discriminative contribution of each feature dimensions, which could facilitate decreasing influence or pruning of redundant, 
noisy or ambiguous feature dimensions. 
This concept can be further extended to more complicated metric tensors with adaptive off-diagonal elements, namely 
Generalized Matrix LVQ (GMLVQ) \citep{schneiderBiehl, petrabiehl}, Limited Rank Matrix LVQ (LiRaM LVQ) \citep{BUNTELiram, kbuntethesis} and 
localized versions with prototype-wise or class-wise matrices called Localized GMLVQ (LGMLVQ) \citep{schneiderBiehl}. 
All algorithms are made publicly available in Matlab and can be found at \url{https://github.com/kbunte/LVQ_toolbox}.

The overlapping class regions as shown in the PCA projection Figure \ref{fig: datadefinition}(a) intuitively suggest non-linear class boundaries 
and hence the localized adaptive metrics are more suitable and therefore the focus for this paper. 
Local metric tensors allow to learn localized dissimilarities with respect to the specific class prototypes using a local transformation matrix $\Omega^j$ 
thus defining the non-linear decision boundaries. 
The localized distance metric is defined as:
\begin{equation}
\label{eq:lgmlvqdist}
d^{\Lambda ^ j} (\vec{\xi}_i, \vec{\omega}^j) = (\vec{\xi}_i-\vec{\omega}^j)^\top \Lambda^{j} (\vec{\xi}_i-\vec{\omega}^j) \enspace,
\end{equation}
where $\Lambda^{j} = \Omega^{j\top}\Omega^j$ using the adaptive matrix $\Omega^j\in\mathbb{R}^{M\times N}$ with $M\le N$ to guarantee that $\Lambda^j$ is positive semi-definite.
The cost function therefore reads as follows:
\begin{equation}
\label{eq:LGMLVQ}
E_\mathrm{LGMLVQ} = \sum_{i=1}^n \Phi(\mu_\mathrm{local}^i) \quad \mathrm{where } \quad \mu_\mathrm{local}^i = \frac{d^{\Lambda^J} - d^{\Lambda^K}}
{d^{\Lambda^J} + d^{\Lambda^K}}
\end{equation}
where $d^{\Lambda^J}$ and $d^{\Lambda^K}$ are the distances of the point $\vec{\xi}_i$ from the closest correct and incorrect prototypes respectively. 
The update rules are described in detail in \cite{petrabiehl, kbuntethesis}. 
Besides allowing non-linear decision boundaries and therefore learning of more complex classification problems, the localized matrices furthermore enable 
the investigation of localized or class-wise relevances, marked on each diagonal of $\Lambda^j$, 
identifying features that are important for the classification of each class respectively \citep{schneiderBiehl}. 

\subsection{Ensemble methods}

With increasing complexity, classifiers get more powerful showing impressive performance in practice. 
However, at the same time they often show overfitting effects in which the performance on training data is near perfect but it decreases facing new data samples not seen before.
This decreased \emph{generalization} performance is often tackled using ensemble methods, which combine several classifiers to assign a class label to a new data instance 
to overcome the limitations of a single model.
In order to see the effect of ensemble learning and to facilitate a fair comparison with RF we explore an ensemble of LGMLVQ models. 
For exact
comparison with RF the ensemble of LGMLVQ models is constructed from the same training bootstrap samples used in each decision tree in the random forest. 
This will result in as many LGMLVQ models as the number of decision trees for each cross validation fold. 
The results are then aggregated through majority voting \citep{ranawana2006}.

\subsection{Interpretability}

For many applications it is crucial for machine learning models to be interpretable, such that the domain expert is able to examine the significance 
of the resulting trained model and its suitability for classification tasks. 
Intrinsic model interpretability can be understood as how understandable the internals of a model and its output are to users \citep{explainableai}. 
It is furthermore suggestively explained by Backhaus and Seiffert \citep{backhaus2014classification} through a three-fold criteria of the model's 
1) feature selection capability, 
2) ability to define class representatives, such as prototypes and 
3) encoding of classification boundary information. 
Interpretability for the random forest is achieved through random permutation of a feature's observations \citep{breiman2001random,strobl2007bias} 
for the out of bag samples and estimating the corresponding decision tree's accuracy with the permuted features. 
Here, more discriminative features are easily identified, since they have significant effect on the classification error. 
The out of bag predictor importance uncovers the individual impact of the features and the information could similarly be used for feature selection and 
understanding the random forest's classification.
On the other hand, the adaptive LVQ methods satisfy all three criteria by: 
1) Feature selection by means of the diagonal of the metric tensors, $\Lambda$ from GMLVQ and the local $\Lambda^j$ in LGMLVQ,
that represents individual feature contribution that could be used as feature pruning criteria. 
2) Prototype feature values used to classify novel observations
are learned during the model training, which subsequently become typical representatives of their corresponding classes. 
3) The decision boundaries for classification 
can be extracted and visualised, for example by linear projection of the data samples and the resulting class prototypes using $\Omega$ from GMLVQ. 
Nonlinear visualizations using the localized variants LiRaMLVQ and LGMLVQ are also possible and the latter is outlined in the following.

\subsection{Nonlinear visualization with charting}\label{sec:charting}

Visualization can be useful to get a holistic view of the data and identify difficult instances.
From the definition of $\Lambda^{j} = \Omega^{j\top}\Omega^j$ in Eq.\ \eqref{eq:lgmlvqdist}, 
we see that the distance metric first transforms data points using the following local linear projections:
\[
 P_j: \vec{\xi} \rightarrow \Omega^{j\top} ( \vec{\xi} - \vec{\omega}^j ) \enspace .
\]
For specific cases $M \in \{ 2, 3 \}$, the projected data points can be visualized, which can be used for discriminant visualization of the data based on the space 
the classification takes place in. 
However, since the localized metric provides several projections for each sample, it is challenging to study the outputs directly.
In order to tackle this problem, \citep{bunte2009nonlinear} proposed a post-processing step which combines local projections using charting \citep{brand2003charting}
to form a global nonlinear embedding of the data:
\[
 \vec{\xi} \rightarrow \sum_j p_j (\vec{\xi}) B_j ( P_j(\vec{\xi})) \enspace .
\]
Here, $B_j(.):\mathbb{R}^M\rightarrow\mathbb{R}^M$ is an affine transformation and
$p_j (\vec{\xi})$ is the responsibility of local transformation $\vec{\omega}^j$ for the data sample $\vec{\xi}$ with $\sum_jp_j(\vec{\xi})=1$. 
More details about how to compute prototypes' responsibilities and affine transformations can be found in 
\citep{bunte2009nonlinear} with the code made available at \url{https://github.com/kbunte/LVQ_toolbox}.
Using this nonlinear embedding we can easily project data to 2 (or 3) dimensions for 
further investigation of the overlapping regions and difficult samples.

%% file: sections/experiments.tex
\section{Experiments and Discussion}
\label{sec:experiments}
This section shows the results and discussion of the experiments conducted with the localized adaptive distance metric LVQ method (LGMLVQ) coupled with presence or 
absence of resampling as a pre-processing step. 
The performance and feature importance is compared with Random Forest (RF). 
The corresponding feature relevances are examined and discussed in comparison with conventional astronomical expectations. 

\subsection{Experimental setup and Evaluation Measures}

The experiments are set up with a $10$-fold cross validation where the data observations are divided up into a $90/10$ random but stratified training and test splits 
with each individual class preserving its sample frequency. 
For the distance based classifiers, such as LGMLVQ, each training set is normalised via Z-score transformation, i.e.\ zero mean and unit standard deviation, 
with the same parameters used to transform the respective test set.
Decision trees and RF are not distance based and build instead rules on the features directly and therefore do not require transformative pre-processing in general.
However, the RF is very sensitive to class imbalance, which should be handled before training.
Therefore, resampling of the training data can be introduced to reduce or eliminate the imbalance of the classes. 
In this contribution we compare two strategies, namely the Synthetic Minority Oversampling Technique (SMOTE) and Borderline-SMOTE, 
creating new feature vectors using the training samples of each minority class until their amount matches the size of the majority class.
The created synthetic minority samples lead to balanced classes to be used for training of the classification models.

The different models have different hyper-parameters.
In the experiments we train the RF with 100 decision trees, sampling with replacement of 0.75 fractions of the training set and 
using the bagging aggregation method \citep{breiman2001random}.
The LVQ models provide several hyper-parameters to control the methods complexity, such as the number of prototypes, 
number of metric tensors and their rank determining the projection dimension saving memory and enabling visualization.
Due to the non-linearity of the problem we use the localized, most powerful version of the LVQ family, namely LGMLVQ \citep{schneider2010regularization}
with one prototype per class and regularization of $10^{-5}$. 
In order to choose the lower dimensional projection dimension we train the method using full metric tensors constructed using $\Omega^j\in\mathbb{R}^{M\times N}$ with $M=N$. 
Subsequently, an Eigenvalue decomposition of the trained $\Lambda^j=U^j\Sigma^jU^{j(-1)}$ with diagonal matrices $\Sigma_{ii}^j=\sigma_i^j$ provides information about the 
intrinsic dimensionality of the classification problem by counting the Eigenvalues $M=\max_{j}\sum_{i=1}^N[\sigma^j_i>\epsilon]$ significantly larger than zero. 
Subsequently, a model is than trained with the rank $M$ reduced to maximal number obtained from the matrices. 
Since the RF is an ensemble of decision trees, we also build an ensemble of LGMLVQ models using the same training sets, and the resulting label for a given sample 
is determined by majority vote among the LGMLVQ models. 
Model performances
are averaged across the cross-validation runs and evaluated with focus on the class of interest, namely UCDs/GCs Class 2. 

For evaluation output statistics are generated after prediction with the models, i.e.\ training and test errors, and their standard deviations. 
Since this is a multi-class classification problem and the major interest is in Class 2 UCD/GC objects, we extract the evaluation measures for each class and build the 
macro averaged accuracies to evaluate the performance across the different classes and eliminate bias of the majority 
class. 
We also report binary class measures, such as Purity and Completeness (also known as Precision/Positive Predictive Value and Recall/Sensitivity),
for the class of interest versus all others classes combined, to represent the algorithms performance in classifying the unseen 
test data.
The confusion matrix as illustrated in Table~\ref{fig:confmatdiag} can be used to evaluate the classification performance providing detailed information about the 
accuracy for each class and the nature of misclassifications.
From it one can extract the binary measures, namely false positives (FP), true positives (TP), false negatives (FN) and true negatives (TN) 
as shown below Table \ref{fig:confmatdiag} in Eq.\ \eqref{eq:binary_problem}.
Additionally, the false positive rate and true positive rate of the test set are plotted on a Receiver Operating Characteristic (ROC) curve \citep{fawcett2006introduction}.
This curve shows the model's discriminative ability and the Area Under the Curve (AUC) summarizes the overall performance for the classification of the class of interest.
\def\myColW{0.12\linewidth}
\begin{table}
\centering
\caption{Three-class confusion matrix and formulae to obtain the False Negatives, True Positives, False Positives and True Negatives 
($FP_b$, $TP_b$, $FP_b$ and $TN_b$) to calculate 
Purity and Completeness 
with respect to class $b$.}
\label{fig:confmatdiag}
\renewcommand{\arraystretch}{1.2}
\begin{tabularx}{0.76\linewidth}{@{\extracolsep{\fill}} 
|c| *{4}{>{\centering\arraybackslash}p{\myColW}|}}
\hline
\diaghead{\theadfont ActualPredicted}{Actual}{Predicted} & Class 1 & Class 2 & Class 3\\\hline
Class 1 & $C_{11}$ & $C_{12}$ & $C_{13}$ \\\cline{2-4}
Class 2 & $C_{21}$ & $C_{22}$ & $C_{23}$ \\\cline{2-4}
Class 3 & $C_{31}$ & $C_{32}$ & $C_{33}$ \\\hline
\end{tabularx}
\renewcommand{\arraystretch}{1}
\begin{align}
\notag
 FN_b &= \textstyle\sum_{\substack{f= 1 \\ f\neq b}}^3 C_{bf} 
&FP_b &= \textstyle\sum_{\substack{f= 1 \\ f\neq b}}^3 C_{fb} \\
\notag
 TN_b &= \textstyle\sum_{\substack{f= 1 \\ f\neq b}}^3 \sum_{\substack{q= 1 \\ q\neq b}}^3 C_{fqx}
&TP_b &= C_{bb} \\
 \mathrm{Purity} &= \textstyle\frac{TP_b}{TP_b + FP_b}
&\mathrm{Compl.} &= \textstyle\frac{TP_b}{TP_b + FN_b}
\label{eq:binary_problem}
\end{align}
\vspace*{-0.85cm}
\end{table}

\subsection{Discusssion}

In this section we summarize the results of the methods on the astronomical classification problem.
The hyperparameter settings for the techniques LGMLVQ (T=L$_M$, with the subscript $M$ indicating the rank for the local metric tensors) 
and Random Forest (T=RF) is abbreviated by $_{\{\varnothing|B|R\}}$T$^{\{\varnothing|\textrm{E}\}}$. 
Here the preprocessing is marked by letters $\mathrm{R}$ and $\mathrm{B}$ for resampling with SMOTE or Borderline-SMOTE or no resampling 
(leaving the prefix empty: $\varnothing$), 
and the superscript $\mathrm{E}$ denotes the results of an ensemble consisting of 100 models.
The performance of the LGMLVQ models averaged across the 10 cross-validation folds is excellent already with an intrinsic dimensionality of $M=2$ 
even without resampling, as evident from the Purity and Completeness 
of the minority class shown in Table \ref{tab:Results}. 
Resampling improves the correct classification 
for the minority class as demonstrated by $3.3\%$ increase in 
the Completeness. 
However, the Purity 
reduces indicating that there are more false positives brought about by SMOTE resampling, which is an acceptable trade-off. 
\begin{table}[t]
\def\myColW2{0.185\linewidth}
\caption{
Average performance (standard deviation) for techniques $_{\{\varnothing|B|R\}}$T$^{\{\varnothing|\textrm{E}\}}$, i.e.\ LGMVLQ (T=L$_M$) and Random Forest (T=RF), 
with no resampling, SMOTE,  Borderline-SMOTE and Ensembling (abbreviated by $\varnothing$, $R$, $B$ and $E$). 
}
\label{tab:Results}
\centering
\begin{tabularx}{\linewidth}{@{\extracolsep{\fill}} @{}>{\footnotesize}l@{} *{5}{@{}>{\footnotesize\centering\arraybackslash}p{\myColW2}@{}} }
                       & \textbf{Purity} & \textbf{Compl.} & \textbf{F1 score} & \textbf{Train accur.} & \textbf{Test accur.} \\ 
\toprule
L$_2$                        & 0.969 (.012) & 0.930 (.021) & 0.947 (.014) & 0.985 (.001) & 0.984 (.004) \\
$_\textrm{R}$L$_\textrm{2}$  & 0.935 (.020) & 0.963 (.020) & 0.948 (.016) & 0.983 (.001) & 0.982 (.005) \\
$_\textrm{B}$L$_\textrm{2}$  & 0.889 (.026) & 0.950 (.027) & 0.912 (.016) & 0.971 (.005) & 0.971 (.006) \\
$_\textrm{R}$L$^\textrm{E}_2$& 0.937 (.019) & 0.962 (.020) & 0.948 (.000) & 0.983 (.001) & 0.983 (.005) \\
$_\textrm{R}$RF$^\textrm{E}$ & 0.950 (.018) & 0.964 (.018) & 0.968 (.011) & 0.999 (.000) & 0.986 (.005) \\ 
\bottomrule
\end{tabularx}
\end{table}

\begin{figure}[t]
\centering
\includegraphics[scale=0.55]{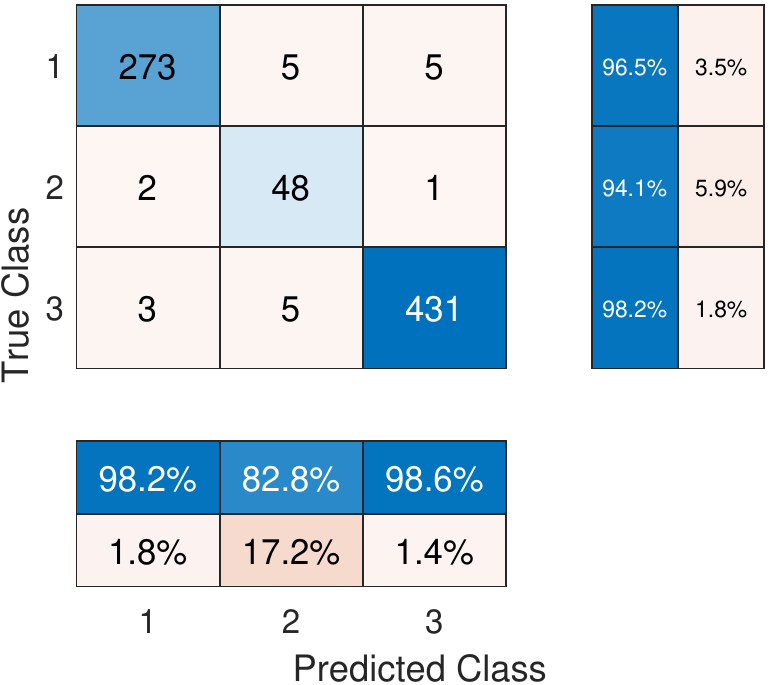}
\hfil
\includegraphics[scale=0.55]{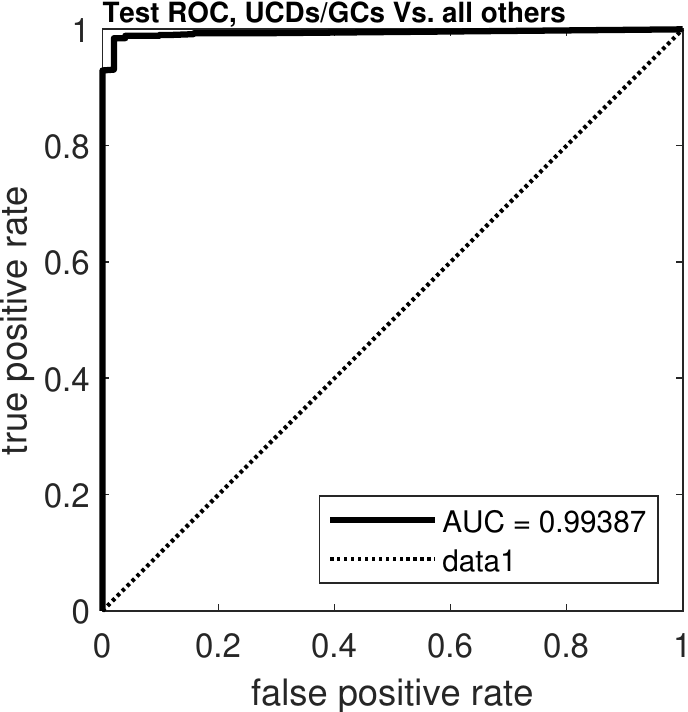}
\caption{Panel (a) shows the test confusion matrix for the $_\mathrm{R}$L$_2$ model
and panel (b) shows the corresponding ROC curve of minority class 2 vs. all the other classes and the corresponding AUC value of $0.99387$.}
\label{fig:misclassifications}
\end{figure}

Figure \ref{fig:misclassifications} panel a) and b) show the test performance of $_\mathrm{R}$L$_\mathrm{2}$ with only $3$ false negatives for the minority class $2$ 
of UCDs and GCs.
Contrary to our expectations Borderline-SMOTE resampling does not perform better. 
This could be caused by the overlap of the classes which increases the difficulty to define a clear boundary and hence boundary resampling becoming ineffective.
In summary the RF and LVQ models show comparable performance. 
However, especially the latter is less complex and provides further insights into the results of the classification, which will be discussed in the following.

\begin{figure}[t]
\centering
\includegraphics[width=\linewidth,trim={0 0.7cm 0 0},clip]{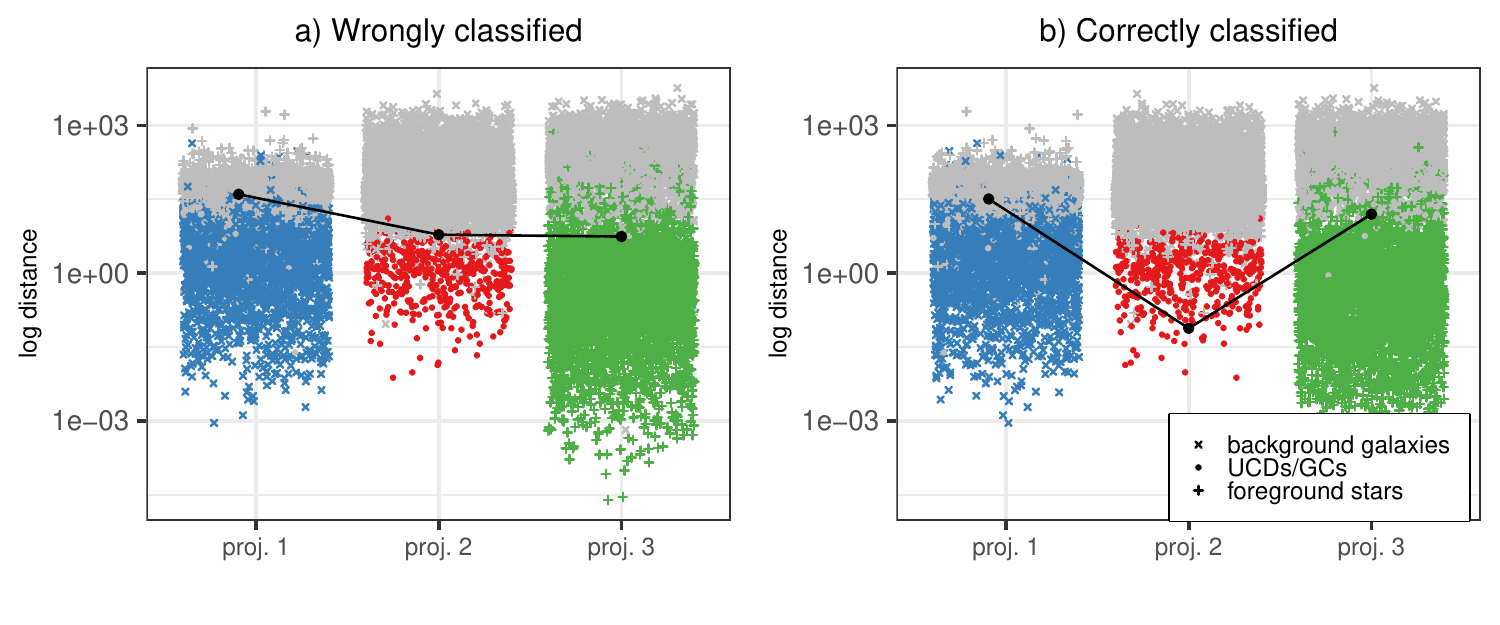}
\caption{Wrong (a) and correct (b) classification of two minority class observations. 
Each shows the distances of all samples projected using the local $\Omega^j$ of each prototype, highlighting the samples with corresponding label $j$ in colour. 
The 
black points exemplify the 
distances of a correct and misclassified sample.}
\label{fig:sampleinter}	
\end{figure}

\begin{figure}[t]
\centering
\includegraphics[width=\linewidth]{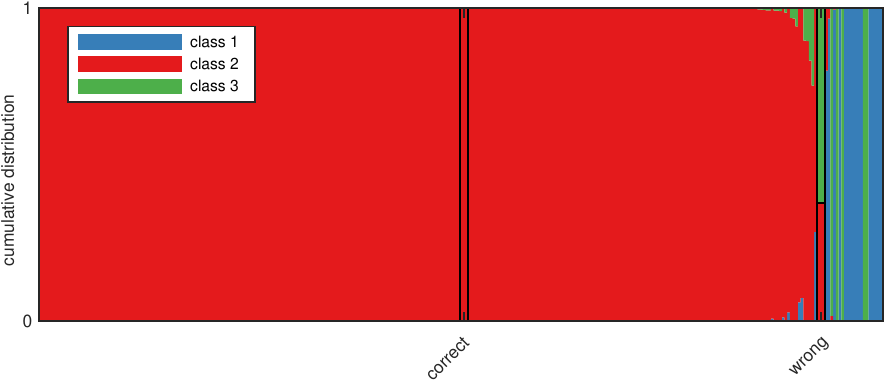}
\caption{Stacked barplot of probabilities $P(j|\vec{\xi}_i)$ (Eq.~\ref{eq:softmax}) of UCDs/GCs samples for each class.
We highlight the probabilities of the same (in)correctly classified examples as specified in Figure \ref{fig:sampleinter}, 
which are [0,1,0] ([0,0.378,0.622]).}
\label{fig:prob}	
\end{figure}

As mentioned before the LVQ models are intrinsically interpretable and transparent in many regards. 
We can for example interpret the certainty of the classification by investigating the distance of each sample to all prototypes.
To demonstrate this we project all samples and all prototypes with the local transformations $\Omega^j$ and compute the distance to each prototype in the transformed space. 
Figure\ \ref{fig:sampleinter} visualizes these distances highlighting the samples with the same respective class label $j$ in colour and of different classes in grey. 
We furthermore highlight in black the distances of an observations consistently misclassified in repeated training runs (panel a) and a correctly classified sample (panel b). 
The wrongly classified minority sample in (a) is within the boundary region where the classes 2 and 3 overlap as indicated by very similar distances to the prototypes of 
these classes. 
Panel (b), on the other hand, shows a typical example of a very certain correct classification where the sample is much closer to prototype 2 compared to the others.
A similar conclusion can be drawn from Figure \ref{fig:prob}. 
It visualizes the probabilities $P(j|\vec{\xi}_i)$ Eq.~\eqref{eq:softmax} as stacked barplot for each minority class sample 
($\vec{\xi}_i$ with $y_i=2$) for each class $j$, and highlights the same samples as depicted in black in Figure\ \ref{fig:sampleinter}. 
While the classifier is certain about the label of the correctly classified example, it is not the case for the misclassified one.
This transparently informs the user how sure the LGMLVQ classifier is with the assignment of a class label for each observation, 
which may also indicate samples with potentially controversial current label identification given the data at hand, marking them as candidates for further investigation.

\begin{figure}[t]
\centering
\begin{tikzpicture}[node distance = 0.1cm,nodes = {anchor=north},>=latex]
\node[inner sep=1pt,text height=3.1cm]             (a)             {\includegraphics[width=0.49\linewidth,trim={0 0.5cm 8.6cm 0},clip]{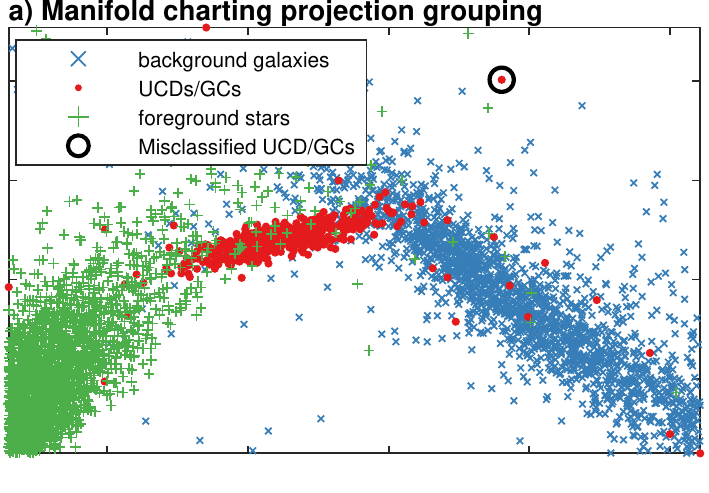}}; 
\node[inner sep=1pt,anchor=west,text height=3.1cm] (b) at (a.east) {\includegraphics[width=0.49\linewidth,trim={7.75cm 0.05cm 0 0},clip]{classifications1}};
\end{tikzpicture}
\caption{a) Manifold charting projection of the data by LGMLVQ ($_\mathrm{R}$L$_{2}$) 
with a misclassified UCD/GC sample circled in black as reference for Figure~\ref{fig:difficultexample}. 
b) conventional color-color diagram used in photometric selection by Astronomy.}
\label{fig:ccvslgmlvq}	
\end{figure}
Moreover, the local linear projections $\Omega^j\in\mathbb{R}^{M\times N}$ can be used for nonlinear visualization for $M\in\{2,3\}$ using manifold charting 
as outlined in section \ref{sec:charting}. 
Hence we report very good performance for LGMLVQ using the rank $M=2$, we show the resulting projected data of $_\mathrm{R}$L$_{2}$ to complement the data analysis 
in Figure\ \ref{fig:ccvslgmlvq} panel a.
The more distinctive separation provided by the LGMLVQ model, especially for the minority class, explains the efficiency and nuance of the method's classification 
performance as compared 
to the traditional astronomical color-color diagram as shown in panel b. 
This visualization shows the difficult regions and can be used to identify difficult observations, such as the circled class 2 sample in panel a, 
in the now reduced overlapping areas.

As mentioned before the RF and LGMLVQ models allow to extract the importance of the measurements for the classification problem.
However, the Random Forest method is expensive in terms of memory costs and shows a clear tendency to overfit as seen in the test error being higher than the training error. 
\begin{figure*}[!ht]
\includegraphics[width=\textwidth]{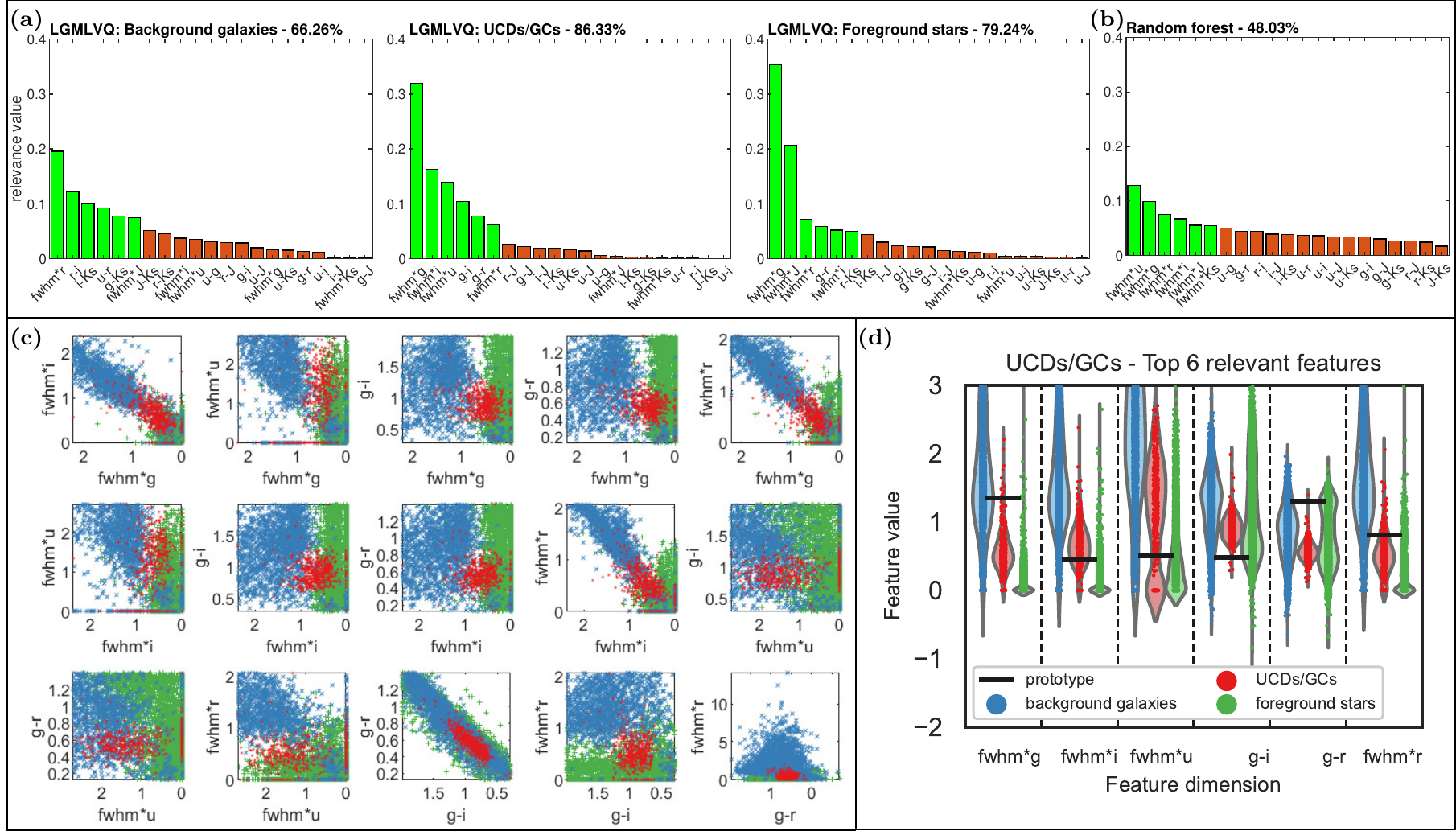}
\caption{Panel (a): class-wise feature relevance profiles of LGMLVQ $_\mathrm{R}$L$_{2}$ marking the top 6 in green and their corresponding percentage 
of contribution to the respective class and (b) the Random Forest (RF) feature relevance profile. 
Panel (c): pairwise plots of the $_\mathrm{R}$L$_{2}$ top 6 features for the UCDs/GCs class 2 (red markers) with focus on the area covered by that class, and 
(d) corresponding violin plots with prototype positions relative to the feature value distribution for minority class $2$.}
\label{fig:lgmlvqrel}
\end{figure*}
Panel (b) in Figure\ \ref{fig:lgmlvqrel} shows the dominance of the angular size features $FWHM^{\ast}u$, $FWHM^{\ast}g$, $FWHM^{\ast}r$, $FWHM^{\ast}i$, $FWHM^{\ast}J$ and $FWHM^{\ast}Ks$ 
in importance for the Random Forest classifier. 
In contrast to RF the LGMLVQ classifier extracts the feature relevances for each prototype and hence we can discuss also the relevance of the measurements for the 
classification of each class of objects in our astronomical data.

Figure \ref{fig:lgmlvqrel}(a) shows the class-wise normalized feature importance profiles sorted by value of contribution with varying top relevant features 
further explaining the non-linearity and motivation for choice of local metric tensors.
The top relevant features for classifying the minority class of UCDs and GCs from this experiment dominantly consist of the angular size parameters, 
namely $FWHM^{\ast}g$, $FWHM^{\ast}i$, $FWHM^{\ast}u$ and $FWHM^{\ast}r$ 
and their pairwise correlation plots are visualized in panel (c) in Figure\ \ref{fig:lgmlvqrel}.
Similarly panel (d) provides additional information in form of a violin plot for the 6 features most important for the classification of class 2, showing the 
distributions of the measurements of each class together with the value the class prototype exhibits after training.
\begin{figure}[!t]
\centering
\includegraphics[width=0.96\linewidth]{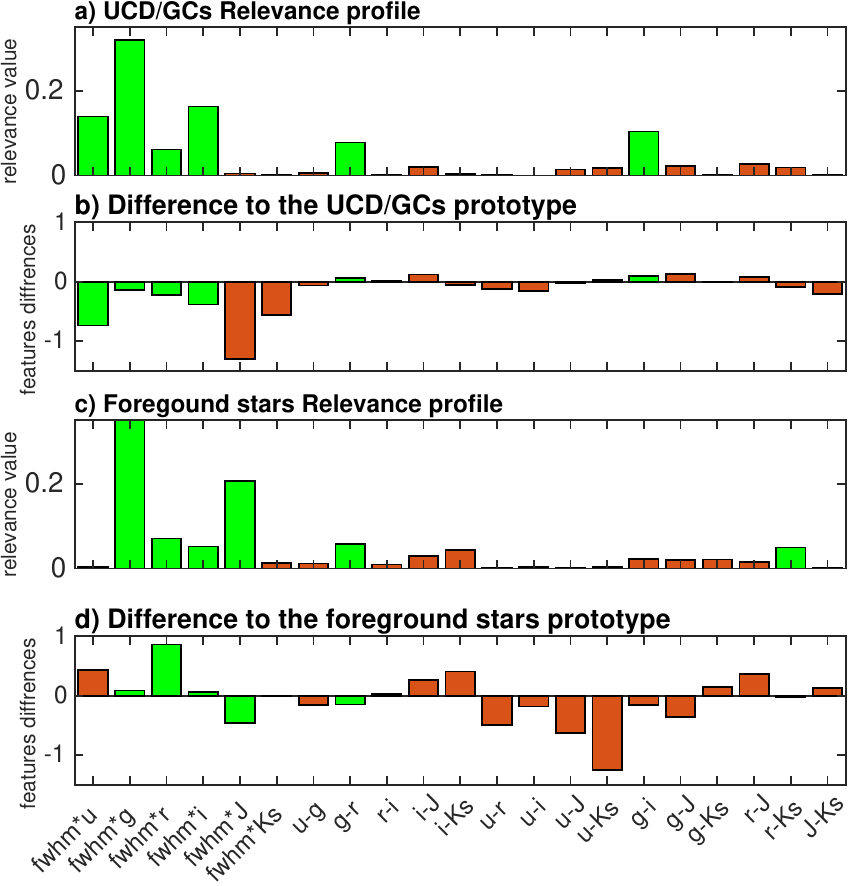}
\caption{
Class-wise relevance profiles of the closest correct and incorrect prototype (panel a and c)
to the difficult UCD/GCs example marked in Figure \ref{fig:ccvslgmlvq}a), with the 6 most relevant features coloured green. 
Panels (b) and (d) show the difference between the sample and 
the prototypes (i.e. $\vec{\xi}_i - \vec{\omega}^J$ and $\vec{\xi}_i - \vec{\omega}^K$). 
}

\label{fig:difficultexample}	
\end{figure}
To examine further the difficult misclassified UCD/GCs example circled in Figure \ref{fig:ccvslgmlvq}a), 
we visualize in Figure \ref{fig:difficultexample}(b) and (d)
the difference between the feature values of the example to the closet correct/incorrect prototypes 
(i.e. $\vec{\xi}_i - \vec{\omega}^J$ and $\vec{\xi}_i - \vec{\omega}^K$).
The 6 most important measurements for the classification are provided by the class-specific relevance profile (a and c) 
on top of each prototype deviation panel and marked in green as before. 
Panels (b) and (d) show that the difference between features of the example to those of the 
foreground stars prototype (for the class-wise most important 6 features) is smaller, which explains the misclassification. 
Such examples require more accurate measurements sizes (using deeper observations) to investigate whether the object indeed belongs to the expected class.

The 
angular size features are important for separating the majority of objects in class $1$, whose objects are larger, from class $2$ and $3$ that have small sizes.
The astronomical expectation is that the angular size cannot discriminate class $2$ and $3$ as illustrated by the model. 
The reason is, that the majority of objects of class $2$ (UCDs/GCs) have sizes smaller than 10 pc. 
Therefore, these objects at the distance of Fornax cluster (20\,Mpc) are expected to appear as point-sources and exhibiting a similar 
FWHM$^*$ value as the objects of class $3$ (foreground stars). 
In this case, FWHM$^*$ would not be an informative parameter in separating class $2$ and $3$ objects.
However, in contrary to astronomical expectation, the angular sizes are found to be important to distinguish class $2$ and $3$ by both the LGMLVQ and RF, 
as shown in Figure\ \ref{fig:lgmlvqrel}. 
The disparity can be attributed to the measurement biases: The minority class $2$ objects are faint (and lower in signal-to-noise ratio) and 
hence angular size measurements happen to be larger than the actual size which could introduce a bias to the data. 
Therefore, their angular size proxies, i.e.\ $FWHM^{\ast}g$, $FWHM^{\ast}i$, $FWHM^{\ast}u$ and $FWHM^{\ast}r$, 
could possess more discriminating information than colour indices.

Based on \citep{munoz2013next}, in the combined optical/near-infrared observations, the color indices of $u-i$ and $i-Ks$ are expected to be the most important features. 
In a simple view, the $u-i$ color is more sensitive to the age of an object while $i-Ks$ represents the metallicity (the amount of metals heavier than Helium) of the object. 
Other color indices, such as $g-i$, $g-r$, $r-J$ etc. could also partially carry these information but degenerate. 
In contrary, the observations in the $u$ and $Ks$ are harder to be carried out and often have a lower signal-to-noise ratio compared to the other filters. 
This makes the expected feature importance of $u-i$ and $i-Ks$ relatively uncertain. 
Values of $g-r$ and $g-i$ on the other hand are notably accurate. 

%% file: sections/conclusion.tex
\section{Conclusion}
\label{sec:conclusion}
In this paper, we explore and compare two interpretable machine learning techniques, namely Localized General Matrix LVQ (LGMLVQ) and Random Forest (RF), 
for the analysis 
and classification of foreground stars and background galaxies versus UCDs and GCs.
Due to the class of interest being highly underrepresented compared the former objects we also investigate the influence of Synthetic Minority Oversampling TEchnique (SMOTE) and 
its borderline extension on the classification performance.
Localized distances allowing non-linear decision boundaries within the data improves the classification in LGMLVQ, 
even in this situation where the classes largely overlap. 
LGMLVQ is also highly interpretable in the form of prototype class representatives and feature relevances which attach values to the contribution of a feature to classification. 
Additionally, the experiments uncover classification patterns in terms of feature relevances, which serve as discriminative markers for the classification. 

The $u-i$ and $i-Ks$ colors are expected to be the most relevant colors for classification since they carry physical information on ages and metallities of astronomical objects. 
However, higher signal-to-noise ratio colors such as $g-i$ and $r-J$ in LGMLVQ and $u-g$ and $g-r$ in Random Forest are found to be  more important for the data-driven classification. 
The importance of other colors compared to $u-i$ and $i-Ks$, that have almost $0$ relevance contribution, implies that this disparity may be attributed to astronomically expected 
features having uncertain measurements, but also the correlation across the features, meaning that they partially contain the same critical information. 
Furthermore, angular size features $FWHM^{\ast}g$, $FWHM{\ast}i$, $FWHM^{\ast}u$ are identified by both methods independently as the most important features for the classification. 
We discuss that this outcome is due to a measurements biases of the faint sources of class 2 (UCDs/GCs).

In this work we showed that existing machine learning techniques can be used to identify UCDs/GCs in big astronomical data. 
These methods can handle the imbalance in the data and classify sources with a good performance. 
Subsequent analysis of the transparent techniques allows further insight and can provide valuable information for the astronomical experts to inform about possible biases in the data set. 
A future deeper data set with more accurate size and color measurements will most likely increase the performance of the automated classification even further.

%% file: journalpaper.bbl
\begin{thebibliography}{43}
\expandafter\ifx\csname natexlab\endcsname\relax\def\natexlab#1{#1}\fi
\providecommand{\url}[1]{\texttt{#1}}
\providecommand{\href}[2]{#2}
\providecommand{\path}[1]{#1}
\providecommand{\DOIprefix}{doi:}
\providecommand{\ArXivprefix}{arXiv:}
\providecommand{\URLprefix}{URL: }
\providecommand{\Pubmedprefix}{pmid:}
\providecommand{\doi}[1]{\href{http://dx.doi.org/#1}{\path{#1}}}
\providecommand{\Pubmed}[1]{\href{pmid:#1}{\path{#1}}}
\providecommand{\bibinfo}[2]{#2}
\ifx\xfnm\relax \def\xfnm[#1]{\unskip,\space#1}\fi
\bibitem[{{Angora} et~al.(2019){Angora}, {Brescia}, {Cavuoti}, {Paolillo},
  {Longo}, {Cantiello}, {Capaccioli}, {D'Abrusco}, {D'Ago}, {Hilker}, {Iodice},
  {Mieske}, {Napolitano}, {Peletier}, {Pota}, {Puzia}, {Riccio} and
  {Spavone}}]{angora}
\bibinfo{author}{{Angora}, G.}, \bibinfo{author}{{Brescia}, M.},
  \bibinfo{author}{{Cavuoti}, S.}, \bibinfo{author}{{Paolillo}, M.},
  \bibinfo{author}{{Longo}, G.}, \bibinfo{author}{{Cantiello}, M.},
  \bibinfo{author}{{Capaccioli}, M.}, \bibinfo{author}{{D'Abrusco}, R.},
  \bibinfo{author}{{D'Ago}, G.}, \bibinfo{author}{{Hilker}, M.},
  \bibinfo{author}{{Iodice}, E.}, \bibinfo{author}{{Mieske}, S.},
  \bibinfo{author}{{Napolitano}, N.}, \bibinfo{author}{{Peletier}, R.},
  \bibinfo{author}{{Pota}, V.}, \bibinfo{author}{{Puzia}, T.},
  \bibinfo{author}{{Riccio}, G.}, \bibinfo{author}{{Spavone}, M.},
  \bibinfo{year}{2019}.
\newblock \bibinfo{title}{{Astroinformatics-based search for globular clusters
  in the Fornax Deep Survey}}.
\newblock \bibinfo{journal}{MNRAS} \bibinfo{volume}{490},
  \bibinfo{pages}{4080--4106}.
\newblock \DOIprefix\doi{10.1093/mnras/stz2801},
  \href{http://arxiv.org/abs/1910.01884}{\tt arXiv:1910.01884}.
\bibitem[{Backhaus and Seiffert(2014)}]{backhaus2014classification}
\bibinfo{author}{Backhaus, A.}, \bibinfo{author}{Seiffert, U.},
  \bibinfo{year}{2014}.
\newblock \bibinfo{title}{Classification in high-dimensional spectral data:
  Accuracy vs. interpretability vs. model size}.
\newblock \bibinfo{journal}{Neurocomputing} \bibinfo{volume}{131},
  \bibinfo{pages}{15--22}.
\bibitem[{Ball et~al.(2004)Ball, Loveday, Fukugita, Nakamura, Okamura,
  Brinkmann and Brunner}]{ball2004galaxy}
\bibinfo{author}{Ball, N.M.}, \bibinfo{author}{Loveday, J.},
  \bibinfo{author}{Fukugita, M.}, \bibinfo{author}{Nakamura, O.},
  \bibinfo{author}{Okamura, S.}, \bibinfo{author}{Brinkmann, J.},
  \bibinfo{author}{Brunner, R.J.}, \bibinfo{year}{2004}.
\newblock \bibinfo{title}{Galaxy types in the sloan digital sky survey using
  supervised artificial neural networks}.
\newblock \bibinfo{journal}{Monthly Notices of the Royal Astronomical Society}
  \bibinfo{volume}{348}, \bibinfo{pages}{1038--1046}.
\bibitem[{Barchi et~al.(2020)Barchi, de~Carvalho, Rosa, Sautter, Soares-Santos,
  Marques, Clua, Gon{\c{c}}alves, de~S{\'a}-Freitas and
  Moura}]{barchi2020machine}
\bibinfo{author}{Barchi, P.}, \bibinfo{author}{de~Carvalho, R.},
  \bibinfo{author}{Rosa, R.}, \bibinfo{author}{Sautter, R.},
  \bibinfo{author}{Soares-Santos, M.}, \bibinfo{author}{Marques, B.},
  \bibinfo{author}{Clua, E.}, \bibinfo{author}{Gon{\c{c}}alves, T.},
  \bibinfo{author}{de~S{\'a}-Freitas, C.}, \bibinfo{author}{Moura, T.},
  \bibinfo{year}{2020}.
\newblock \bibinfo{title}{Machine and deep learning applied to galaxy
  morphology-a comparative study}.
\newblock \bibinfo{journal}{Astronomy and Computing} \bibinfo{volume}{30},
  \bibinfo{pages}{100334}.
\bibitem[{{Beasley}(2020)}]{beasley2020}
\bibinfo{author}{{Beasley}, M.A.}, \bibinfo{year}{2020}.
\newblock \bibinfo{title}{{Globular Cluster Systems and Galaxy Formation}}, in:
  \bibinfo{booktitle}{Reviews in Frontiers of Modern Astrophysics; From Space
  Debris to Cosmology}, pp. \bibinfo{pages}{245--277}.
\newblock \DOIprefix\doi{10.1007/978-3-030-38509-5_9}.
\bibitem[{Brand(2003)}]{brand2003charting}
\bibinfo{author}{Brand, M.}, \bibinfo{year}{2003}.
\newblock \bibinfo{title}{Charting a manifold}, in:
  \bibinfo{booktitle}{Advances in neural information processing systems}, pp.
  \bibinfo{pages}{985--992}.
\bibitem[{Breiman(2001)}]{breiman2001random}
\bibinfo{author}{Breiman, L.}, \bibinfo{year}{2001}.
\newblock \bibinfo{title}{Random forests}.
\newblock \bibinfo{journal}{Machine learning} \bibinfo{volume}{45},
  \bibinfo{pages}{5--32}.
\bibitem[{Bunte(2011)}]{kbuntethesis}
\bibinfo{author}{Bunte, K.}, \bibinfo{year}{2011}.
\newblock \bibinfo{title}{Adaptive dissimilarity measures, dimension reduction
  and visualization}.
\newblock Ph.D. thesis. University of Groningen.
\bibitem[{Bunte et~al.(2009)Bunte, Hammer, Schneider and
  Biehl}]{bunte2009nonlinear}
\bibinfo{author}{Bunte, K.}, \bibinfo{author}{Hammer, B.},
  \bibinfo{author}{Schneider, P.}, \bibinfo{author}{Biehl, M.},
  \bibinfo{year}{2009}.
\newblock \bibinfo{title}{Nonlinear discriminative data visualization.}, in:
  \bibinfo{booktitle}{ESANN}, pp. \bibinfo{pages}{65--70}.
\bibitem[{Bunte et~al.(2012)Bunte, Schneider, Hammer, Schleif, Villmann and
  Biehl}]{BUNTELiram}
\bibinfo{author}{Bunte, K.}, \bibinfo{author}{Schneider, P.},
  \bibinfo{author}{Hammer, B.}, \bibinfo{author}{Schleif, F.M.},
  \bibinfo{author}{Villmann, T.}, \bibinfo{author}{Biehl, M.},
  \bibinfo{year}{2012}.
\newblock \bibinfo{title}{Limited rank matrix learning, discriminative
  dimension reduction and visualization}.
\newblock \bibinfo{journal}{Neural Networks} \bibinfo{volume}{26},
  \bibinfo{pages}{159 -- 173}.
\newblock \URLprefix
  \url{http://www.sciencedirect.com/science/article/pii/S0893608011002632}.
\bibitem[{{Cantiello} et~al.(2018){Cantiello}, {Grado}, {Rejkuba}, {Arnaboldi},
  {Capaccioli}, {Greggio}, {Iodice} and {Limatola}}]{Cantiello2018}
\bibinfo{author}{{Cantiello}, M.}, \bibinfo{author}{{Grado}, A.},
  \bibinfo{author}{{Rejkuba}, M.}, \bibinfo{author}{{Arnaboldi}, M.},
  \bibinfo{author}{{Capaccioli}, M.}, \bibinfo{author}{{Greggio}, L.},
  \bibinfo{author}{{Iodice}, E.}, \bibinfo{author}{{Limatola}, L.},
  \bibinfo{year}{2018}.
\newblock \bibinfo{title}{{A VST and VISTA study of globular clusters in NGC
  253}}.
\newblock \bibinfo{journal}{A\&A} \bibinfo{volume}{611}, \bibinfo{pages}{A21}.
\newblock \DOIprefix\doi{10.1051/0004-6361/201731325},
  \href{http://arxiv.org/abs/1711.00805}{\tt arXiv:1711.00805}.
\bibitem[{{Cantiello} et~al.(2020){Cantiello}, {Venhola}, {Grado}, {Paolillo},
  {D'Abrusco}, {Raimondo}, {Quintini}, {Hilker}, {Mieske}, {Tortora},
  {Spavone}, {Capaccioli}, {Iodice}, {Peletier}, {Falcon Barroso}, {Limatola},
  {Napolitano}, {Schipani}, {van de Ven}, {Gentile} and
  {Covone}}]{Cantiello2020}
\bibinfo{author}{{Cantiello}, M.}, \bibinfo{author}{{Venhola}, A.},
  \bibinfo{author}{{Grado}, A.}, \bibinfo{author}{{Paolillo}, M.},
  \bibinfo{author}{{D'Abrusco}, R.}, \bibinfo{author}{{Raimondo}, G.},
  \bibinfo{author}{{Quintini}, M.}, \bibinfo{author}{{Hilker}, M.},
  \bibinfo{author}{{Mieske}, S.}, \bibinfo{author}{{Tortora}, C.},
  \bibinfo{author}{{Spavone}, M.}, \bibinfo{author}{{Capaccioli}, M.},
  \bibinfo{author}{{Iodice}, E.}, \bibinfo{author}{{Peletier}, R.},
  \bibinfo{author}{{Falcon Barroso}, J.}, \bibinfo{author}{{Limatola}, L.},
  \bibinfo{author}{{Napolitano}, N.}, \bibinfo{author}{{Schipani}, P.},
  \bibinfo{author}{{van de Ven}, G.}, \bibinfo{author}{{Gentile}, F.},
  \bibinfo{author}{{Covone}, G.}, \bibinfo{year}{2020}.
\newblock \bibinfo{title}{{The Fornax Deep Survey with VST. X. The catalog of
  sources in the FDS area, with an example study for globular clusters and
  background galaxies}}.
\newblock \bibinfo{journal}{arXiv e-prints} ,
  \bibinfo{pages}{arXiv:2005.12085}\href{http://arxiv.org/abs/2005.12085}{\tt
  arXiv:2005.12085}.
\bibitem[{Carrasco et~al.(2015)Carrasco, Barrientos, Pichara, Anguita, Murphy,
  Gilbank, Gladders, Yee, Hsieh and L{\'o}pez}]{carrasco2015photometric}
\bibinfo{author}{Carrasco, D.}, \bibinfo{author}{Barrientos, L.},
  \bibinfo{author}{Pichara, K.}, \bibinfo{author}{Anguita, T.},
  \bibinfo{author}{Murphy, D.N.}, \bibinfo{author}{Gilbank, D.G.},
  \bibinfo{author}{Gladders, M.D.}, \bibinfo{author}{Yee, H.K.},
  \bibinfo{author}{Hsieh, B.C.}, \bibinfo{author}{L{\'o}pez, S.},
  \bibinfo{year}{2015}.
\newblock \bibinfo{title}{Photometric classification of quasars from rcs-2
  using random forest}.
\newblock \bibinfo{journal}{Astronomy \& Astrophysics} \bibinfo{volume}{584},
  \bibinfo{pages}{A44}.
\bibitem[{Chawla et~al.(2002)Chawla, Bowyer, Hall and Kegelmeyer}]{smotechawla}
\bibinfo{author}{Chawla, N.V.}, \bibinfo{author}{Bowyer, K.W.},
  \bibinfo{author}{Hall, L.O.}, \bibinfo{author}{Kegelmeyer, W.P.},
  \bibinfo{year}{2002}.
\newblock \bibinfo{title}{Smote: Synthetic minority over-sampling technique}.
\newblock \bibinfo{journal}{J. Artif. Int. Res.} \bibinfo{volume}{16},
  \bibinfo{pages}{321–357}.
\bibitem[{{D'Abrusco} et~al.(2016){D'Abrusco}, {Cantiello}, {Paolillo}, {Pota},
  {Napolitano}, {Limatola}, {Spavone}, {Grado}, {Iodice}, {Capaccioli},
  {Peletier}, {Longo}, {Hilker}, {Mieske}, {Grebel}, {Lisker}, {Wittmann}, {van
  de Ven}, {Schipani} and {Fabbiano}}]{DAbrusco}
\bibinfo{author}{{D'Abrusco}, R.}, \bibinfo{author}{{Cantiello}, M.},
  \bibinfo{author}{{Paolillo}, M.}, \bibinfo{author}{{Pota}, V.},
  \bibinfo{author}{{Napolitano}, N.R.}, \bibinfo{author}{{Limatola}, L.},
  \bibinfo{author}{{Spavone}, M.}, \bibinfo{author}{{Grado}, A.},
  \bibinfo{author}{{Iodice}, E.}, \bibinfo{author}{{Capaccioli}, M.},
  \bibinfo{author}{{Peletier}, R.}, \bibinfo{author}{{Longo}, G.},
  \bibinfo{author}{{Hilker}, M.}, \bibinfo{author}{{Mieske}, S.},
  \bibinfo{author}{{Grebel}, E.K.}, \bibinfo{author}{{Lisker}, T.},
  \bibinfo{author}{{Wittmann}, C.}, \bibinfo{author}{{van de Ven}, G.},
  \bibinfo{author}{{Schipani}, P.}, \bibinfo{author}{{Fabbiano}, G.},
  \bibinfo{year}{2016}.
\newblock \bibinfo{title}{{The Extended Spatial Distribution of Globular
  Clusters in the Core of the Fornax Cluster}}.
\newblock \bibinfo{journal}{ApJ Letters} \bibinfo{volume}{819},
  \bibinfo{pages}{L31}.
\newblock \DOIprefix\doi{10.3847/2041-8205/819/2/L31},
  \href{http://arxiv.org/abs/1602.06076}{\tt arXiv:1602.06076}.
\bibitem[{Delli~Veneri et~al.(2019)Delli~Veneri, Cavuoti, Brescia, Longo and
  Riccio}]{delli2019star}
\bibinfo{author}{Delli~Veneri, M.}, \bibinfo{author}{Cavuoti, S.},
  \bibinfo{author}{Brescia, M.}, \bibinfo{author}{Longo, G.},
  \bibinfo{author}{Riccio, G.}, \bibinfo{year}{2019}.
\newblock \bibinfo{title}{Star formation rates for photometric samples of
  galaxies using machine learning methods}.
\newblock \bibinfo{journal}{Monthly Notices of the Royal Astronomical Society}
  \bibinfo{volume}{486}, \bibinfo{pages}{1377--1391}.
\bibitem[{Fawcett(2006)}]{fawcett2006introduction}
\bibinfo{author}{Fawcett, T.}, \bibinfo{year}{2006}.
\newblock \bibinfo{title}{An introduction to roc analysis}.
\newblock \bibinfo{journal}{Pattern recognition letters} \bibinfo{volume}{27},
  \bibinfo{pages}{861--874}.
\bibitem[{Gao et~al.(2009)Gao, Zhang and Zhao}]{gao2009random}
\bibinfo{author}{Gao, D.}, \bibinfo{author}{Zhang, Y.X.},
  \bibinfo{author}{Zhao, Y.H.}, \bibinfo{year}{2009}.
\newblock \bibinfo{title}{Random forest algorithm for classification of
  multiwavelength data}.
\newblock \bibinfo{journal}{Research in Astronomy and Astrophysics}
  \bibinfo{volume}{9}, \bibinfo{pages}{220}.
\bibitem[{Gilpin et~al.(2018)Gilpin, Bau, Yuan, Bajwa, Specter and
  Kagal}]{explainableai}
\bibinfo{author}{Gilpin, L.H.}, \bibinfo{author}{Bau, D.},
  \bibinfo{author}{Yuan, B.Z.}, \bibinfo{author}{Bajwa, A.},
  \bibinfo{author}{Specter, M.}, \bibinfo{author}{Kagal, L.},
  \bibinfo{year}{2018}.
\newblock \bibinfo{title}{Explaining explanations: An approach to evaluating
  interpretability of machine learning}.
\newblock \bibinfo{journal}{CoRR} \URLprefix
  \url{http://arxiv.org/abs/1806.00069}.
\bibitem[{Hammer and Villmann(2002)}]{hammer2002generalized}
\bibinfo{author}{Hammer, B.}, \bibinfo{author}{Villmann, T.},
  \bibinfo{year}{2002}.
\newblock \bibinfo{title}{Generalized relevance learning vector quantization}.
\newblock \bibinfo{journal}{Neural Networks} \bibinfo{volume}{15},
  \bibinfo{pages}{1059--1068}.
\bibitem[{Han et~al.(2005)Han, Wang and Mao}]{borderlinesmote}
\bibinfo{author}{Han, H.}, \bibinfo{author}{Wang, W.Y.}, \bibinfo{author}{Mao,
  B.H.}, \bibinfo{year}{2005}.
\newblock \bibinfo{title}{Borderline-smote: A new over-sampling method in
  imbalanced data sets learning}, in: \bibinfo{editor}{Huang, D.S.},
  \bibinfo{editor}{Zhang, X.P.}, \bibinfo{editor}{Huang, G.B.} (Eds.),
  \bibinfo{booktitle}{Advances in Intelligent Computing},
  \bibinfo{publisher}{Springer Berlin Heidelberg}, \bibinfo{address}{Berlin,
  Heidelberg}. pp. \bibinfo{pages}{878--887}.
\bibitem[{{Hubble}(1929)}]{hubble}
\bibinfo{author}{{Hubble}, E.}, \bibinfo{year}{1929}.
\newblock \bibinfo{title}{{A Relation between Distance and Radial Velocity
  among Extra-Galactic Nebulae}}.
\newblock \bibinfo{journal}{Proceedings of the National Academy of Science}
  \bibinfo{volume}{15}, \bibinfo{pages}{168--173}.
\newblock \DOIprefix\doi{10.1073/pnas.15.3.168}.
\bibitem[{Jones and Singal(2017)}]{Jones_2017}
\bibinfo{author}{Jones, E.}, \bibinfo{author}{Singal, J.},
  \bibinfo{year}{2017}.
\newblock \bibinfo{title}{Analysis of a custom support vector machine for
  photometric redshift estimation and the inclusion of galaxy shape
  information}.
\newblock \bibinfo{journal}{Astronomy \& Astrophysics} \bibinfo{volume}{600},
  \bibinfo{pages}{A113}.
\newblock \URLprefix \url{http://dx.doi.org/10.1051/0004-6361/201629558}.
\bibitem[{{Jord{\'a}n} et~al.(2009){Jord{\'a}n}, {Peng}, {Blakeslee},
  {C{\^o}t{\'e}}, {Eyheramendy}, {Ferrarese}, {Mei}, {Tonry} and
  {West}}]{jordan2009}
\bibinfo{author}{{Jord{\'a}n}, A.}, \bibinfo{author}{{Peng}, E.W.},
  \bibinfo{author}{{Blakeslee}, J.P.}, \bibinfo{author}{{C{\^o}t{\'e}}, P.},
  \bibinfo{author}{{Eyheramendy}, S.}, \bibinfo{author}{{Ferrarese}, L.},
  \bibinfo{author}{{Mei}, S.}, \bibinfo{author}{{Tonry}, J.L.},
  \bibinfo{author}{{West}, M.J.}, \bibinfo{year}{2009}.
\newblock \bibinfo{title}{{The ACS Virgo Cluster Survey XVI. Selection
  Procedure and Catalogs of Globular Cluster Candidates}}.
\newblock \bibinfo{journal}{The Astrophysical Journal Supplement Series}
  \bibinfo{volume}{180}, \bibinfo{pages}{54--66}.
\newblock \DOIprefix\doi{10.1088/0067-0049/180/1/54}.
\bibitem[{{Lema{\^\i}tre}(1931)}]{lemaitre}
\bibinfo{author}{{Lema{\^\i}tre}, G.}, \bibinfo{year}{1931}.
\newblock \bibinfo{title}{{Expansion of the universe, A homogeneous universe of
  constant mass and increasing radius accounting for the radial velocity of
  extra-galactic nebulae}}.
\newblock \bibinfo{journal}{MNRAS} \bibinfo{volume}{91},
  \bibinfo{pages}{483--490}.
\newblock \DOIprefix\doi{10.1093/mnras/91.5.483}.
\bibitem[{Li et~al.(2008)Li, Zhang and Zhao}]{li2008k}
\bibinfo{author}{Li, L.}, \bibinfo{author}{Zhang, Y.}, \bibinfo{author}{Zhao,
  Y.}, \bibinfo{year}{2008}.
\newblock \bibinfo{title}{k-nearest neighbors for automated classification of
  celestial objects}.
\newblock \bibinfo{journal}{Science in China Series G: Physics, Mechanics and
  Astronomy} \bibinfo{volume}{51}, \bibinfo{pages}{916--922}.
\bibitem[{{McMahon} et~al.(2013){McMahon}, {Banerji}, {Gonzalez}, {Koposov},
  {Bejar}, {Lodieu}, {Rebolo} and {VHS Collaboration}}]{vhs}
\bibinfo{author}{{McMahon}, R.G.}, \bibinfo{author}{{Banerji}, M.},
  \bibinfo{author}{{Gonzalez}, E.}, \bibinfo{author}{{Koposov}, S.E.},
  \bibinfo{author}{{Bejar}, V.J.}, \bibinfo{author}{{Lodieu}, N.},
  \bibinfo{author}{{Rebolo}, R.}, \bibinfo{author}{{VHS Collaboration}},
  \bibinfo{year}{2013}.
\newblock \bibinfo{title}{{First Scientific Results from the VISTA Hemisphere
  Survey (VHS)}}.
\newblock \bibinfo{journal}{The Messenger} \bibinfo{volume}{154},
  \bibinfo{pages}{35--37}.
\bibitem[{{Mo} et~al.(2010){Mo}, {van den Bosch} and {White}}]{galaxyformation}
\bibinfo{author}{{Mo}, H.}, \bibinfo{author}{{van den Bosch}, F.C.},
  \bibinfo{author}{{White}, S.}, \bibinfo{year}{2010}.
\newblock \bibinfo{title}{{Galaxy Formation and Evolution}}.
\bibitem[{Mohammadi et~al.(2019)Mohammadi, Petkov, Bunte, Peletier and
  Schleif}]{mohammadi2019globular}
\bibinfo{author}{Mohammadi, M.}, \bibinfo{author}{Petkov, N.},
  \bibinfo{author}{Bunte, K.}, \bibinfo{author}{Peletier, R.F.},
  \bibinfo{author}{Schleif, F.M.}, \bibinfo{year}{2019}.
\newblock \bibinfo{title}{Globular cluster detection in the gaia survey}.
\newblock \bibinfo{journal}{Neurocomputing} \bibinfo{volume}{342},
  \bibinfo{pages}{164--171}.
\bibitem[{{Mu{\~n}oz} et~al.(2014){Mu{\~n}oz}, {Puzia}, {Lan{\c{c}}on}, {Peng},
  {C{\^o}t{\'e}}, {Ferrarese}, {Blakeslee}, {Mei}, {Cuillandre}, {Hudelot},
  {Courteau}, {Duc}, {Balogh}, {Boselli}, {Bournaud}, {Carlberg}, {Chapman},
  {Durrell}, {Eigenthaler}, {Emsellem}, {Gavazzi}, {Gwyn}, {Huertas-Company},
  {Ilbert}, {Jord{\'a}n}, {L{\"a}sker}, {Licitra}, {Liu}, {MacArthur},
  {McConnachie}, {McCracken}, {Mellier}, {Peng}, {Raichoor}, {Taylor}, {Tonry},
  {Tully} and {Zhang}}]{munoz}
\bibinfo{author}{{Mu{\~n}oz}, R.P.}, \bibinfo{author}{{Puzia}, T.H.},
  \bibinfo{author}{{Lan{\c{c}}on}, A.}, \bibinfo{author}{{Peng}, E.W.},
  \bibinfo{author}{{C{\^o}t{\'e}}, P.}, \bibinfo{author}{{Ferrarese}, L.},
  \bibinfo{author}{{Blakeslee}, J.P.}, \bibinfo{author}{{Mei}, S.},
  \bibinfo{author}{{Cuillandre}, J.C.}, \bibinfo{author}{{Hudelot}, P.},
  \bibinfo{author}{{Courteau}, S.}, \bibinfo{author}{{Duc}, P.A.},
  \bibinfo{author}{{Balogh}, M.L.}, \bibinfo{author}{{Boselli}, A.},
  \bibinfo{author}{{Bournaud}, F.}, \bibinfo{author}{{Carlberg}, R.G.},
  \bibinfo{author}{{Chapman}, S.C.}, \bibinfo{author}{{Durrell}, P.},
  \bibinfo{author}{{Eigenthaler}, P.}, \bibinfo{author}{{Emsellem}, E.},
  \bibinfo{author}{{Gavazzi}, G.}, \bibinfo{author}{{Gwyn}, S.},
  \bibinfo{author}{{Huertas-Company}, M.}, \bibinfo{author}{{Ilbert}, O.},
  \bibinfo{author}{{Jord{\'a}n}, A.}, \bibinfo{author}{{L{\"a}sker}, R.},
  \bibinfo{author}{{Licitra}, R.}, \bibinfo{author}{{Liu}, C.},
  \bibinfo{author}{{MacArthur}, L.}, \bibinfo{author}{{McConnachie}, A.},
  \bibinfo{author}{{McCracken}, H.J.}, \bibinfo{author}{{Mellier}, Y.},
  \bibinfo{author}{{Peng}, C.Y.}, \bibinfo{author}{{Raichoor}, A.},
  \bibinfo{author}{{Taylor}, M.A.}, \bibinfo{author}{{Tonry}, J.L.},
  \bibinfo{author}{{Tully}, R.B.}, \bibinfo{author}{{Zhang}, H.},
  \bibinfo{year}{2014}.
\newblock \bibinfo{title}{{The Next Generation Virgo Cluster Survey-Infrared
  (NGVS-IR). I. A New Near-Ultraviolet, Optical, and Near-Infrared Globular
  Cluster Selection Tool}}.
\newblock \bibinfo{journal}{The Astrophysical Journal Supplement Series}
  \bibinfo{volume}{210}, \bibinfo{pages}{4}.
\newblock \DOIprefix\doi{10.1088/0067-0049/210/1/4},
  \href{http://arxiv.org/abs/1311.0873}{\tt arXiv:1311.0873}.
\bibitem[{Munoz et~al.(2013)Munoz, Puzia, Lan{\c{c}}on, Peng, Cote, Ferrarese,
  Blakeslee, Mei, Cuillandre, Hudelot et~al.}]{munoz2013next}
\bibinfo{author}{Munoz, R.P.}, \bibinfo{author}{Puzia, T.H.},
  \bibinfo{author}{Lan{\c{c}}on, A.}, \bibinfo{author}{Peng, E.W.},
  \bibinfo{author}{Cote, P.}, \bibinfo{author}{Ferrarese, L.},
  \bibinfo{author}{Blakeslee, J.P.}, \bibinfo{author}{Mei, S.},
  \bibinfo{author}{Cuillandre, J.C.}, \bibinfo{author}{Hudelot, P.}, et~al.,
  \bibinfo{year}{2013}.
\newblock \bibinfo{title}{The next generation virgo cluster survey-infrared
  (ngvs-ir). i. a new near-ultraviolet, optical, and near-infrared globular
  cluster selection tool}.
\newblock \bibinfo{journal}{The Astrophysical Journal Supplement Series}
  \bibinfo{volume}{210}, \bibinfo{pages}{4}.
\bibitem[{Nevin et~al.(2019)Nevin, Blecha, Comerford and
  Greene}]{nevin2019accurate}
\bibinfo{author}{Nevin, R.}, \bibinfo{author}{Blecha, L.},
  \bibinfo{author}{Comerford, J.}, \bibinfo{author}{Greene, J.},
  \bibinfo{year}{2019}.
\newblock \bibinfo{title}{Accurate identification of galaxy mergers with
  imaging}.
\newblock \bibinfo{journal}{The Astrophysical Journal} \bibinfo{volume}{872},
  \bibinfo{pages}{76}.
\bibitem[{{Pota} et~al.(2018){Pota}, {Napolitano}, {Hilker}, {Spavone},
  {Schulz}, {Cantiello}, {Tortora}, {Iodice}, {Paolillo}, {D'Abrusco},
  {Capaccioli}, {Puzia}, {Peletier}, {Romanowsky}, {van de Ven}, {Spiniello},
  {Norris}, {Lisker}, {Munoz}, {Schipani}, {Eigenthaler}, {Taylor},
  {S{\'a}nchez-Janssen} and {Ordenes-Brice{\~n}o}}]{pota}
\bibinfo{author}{{Pota}, V.}, \bibinfo{author}{{Napolitano}, N.R.},
  \bibinfo{author}{{Hilker}, M.}, \bibinfo{author}{{Spavone}, M.},
  \bibinfo{author}{{Schulz}, C.}, \bibinfo{author}{{Cantiello}, M.},
  \bibinfo{author}{{Tortora}, C.}, \bibinfo{author}{{Iodice}, E.},
  \bibinfo{author}{{Paolillo}, M.}, \bibinfo{author}{{D'Abrusco}, R.},
  \bibinfo{author}{{Capaccioli}, M.}, \bibinfo{author}{{Puzia}, T.},
  \bibinfo{author}{{Peletier}, R.F.}, \bibinfo{author}{{Romanowsky}, A.J.},
  \bibinfo{author}{{van de Ven}, G.}, \bibinfo{author}{{Spiniello}, C.},
  \bibinfo{author}{{Norris}, M.}, \bibinfo{author}{{Lisker}, T.},
  \bibinfo{author}{{Munoz}, R.}, \bibinfo{author}{{Schipani}, P.},
  \bibinfo{author}{{Eigenthaler}, P.}, \bibinfo{author}{{Taylor}, M.A.},
  \bibinfo{author}{{S{\'a}nchez-Janssen}, R.},
  \bibinfo{author}{{Ordenes-Brice{\~n}o}, Y.}, \bibinfo{year}{2018}.
\newblock \bibinfo{title}{{The Fornax Cluster VLT Spectroscopic Survey - I.
  VIMOS spectroscopy of compact stellar systems in the Fornax core region}}.
\newblock \bibinfo{journal}{MNRAS} \bibinfo{volume}{481},
  \bibinfo{pages}{1744--1756}.
\newblock \DOIprefix\doi{10.1093/mnras/sty2149},
  \href{http://arxiv.org/abs/1803.03275}{\tt arXiv:1803.03275}.
\bibitem[{{Prole} et~al.(2019){Prole}, {Hilker}, {van der Burg}, {Cantiello},
  {Venhola}, {Iodice}, {van de Ven}, {Wittmann}, {Peletier}, {Mieske},
  {Capaccioli}, {Napolitano}, {Paolillo}, {Spavone} and
  {Valentijn}}]{Prole-2019}
\bibinfo{author}{{Prole}, D.J.}, \bibinfo{author}{{Hilker}, M.},
  \bibinfo{author}{{van der Burg}, R.F.J.}, \bibinfo{author}{{Cantiello}, M.},
  \bibinfo{author}{{Venhola}, A.}, \bibinfo{author}{{Iodice}, E.},
  \bibinfo{author}{{van de Ven}, G.}, \bibinfo{author}{{Wittmann}, C.},
  \bibinfo{author}{{Peletier}, R.F.}, \bibinfo{author}{{Mieske}, S.},
  \bibinfo{author}{{Capaccioli}, M.}, \bibinfo{author}{{Napolitano}, N.R.},
  \bibinfo{author}{{Paolillo}, M.}, \bibinfo{author}{{Spavone}, M.},
  \bibinfo{author}{{Valentijn}, E.}, \bibinfo{year}{2019}.
\newblock \bibinfo{title}{{Halo mass estimates from the globular cluster
  populations of 175 low surface brightness galaxies in the Fornax cluster}}.
\newblock \bibinfo{journal}{MNRAS} \bibinfo{volume}{484},
  \bibinfo{pages}{4865--4880}.
\newblock \DOIprefix\doi{10.1093/mnras/stz326},
  \href{http://arxiv.org/abs/1901.09648}{\tt arXiv:1901.09648}.
\bibitem[{Ranawana and Palade(2006)}]{ranawana2006}
\bibinfo{author}{Ranawana, R.}, \bibinfo{author}{Palade, V.},
  \bibinfo{year}{2006}.
\newblock \bibinfo{title}{Multi-classifier systems: Review and a roadmap for
  developers}.
\newblock \bibinfo{journal}{Int. J. Hybrid Intell. Syst.} \bibinfo{volume}{3},
  \bibinfo{pages}{35–61}.
\bibitem[{{Saifollahi} et~al.(2021){Saifollahi}, {Janz}, {Peletier},
  {Cantiello}, {Hilker}, {Mieske}, {Valentijn}, {Venhola} and
  {Kleijn}}]{teymoor}
\bibinfo{author}{{Saifollahi}, T.}, \bibinfo{author}{{Janz}, J.},
  \bibinfo{author}{{Peletier}, R.F.}, \bibinfo{author}{{Cantiello}, M.},
  \bibinfo{author}{{Hilker}, M.}, \bibinfo{author}{{Mieske}, S.},
  \bibinfo{author}{{Valentijn}, E.A.}, \bibinfo{author}{{Venhola}, A.},
  \bibinfo{author}{{Kleijn}, G.V.}, \bibinfo{year}{2021}.
\newblock \bibinfo{title}{{Ultra-compact dwarfs beyond the centre of the Fornax
  galaxy cluster: hints of UCD formation in low-density environments}}.
\newblock \bibinfo{journal}{MNRAS} \bibinfo{volume}{504},
  \bibinfo{pages}{3580--3609}.
\newblock \DOIprefix\doi{10.1093/mnras/stab1118},
  \href{http://arxiv.org/abs/2104.00004}{\tt arXiv:2104.00004}.
\bibitem[{Sato and Yamada(1995)}]{sato&yamada}
\bibinfo{author}{Sato, A.}, \bibinfo{author}{Yamada, K.}, \bibinfo{year}{1995}.
\newblock \bibinfo{title}{Generalized learning vector quantization}, in:
  \bibinfo{booktitle}{Proceedings of the 8th International Conference on Neural
  Information Processing Systems}, \bibinfo{publisher}{MIT Press},
  \bibinfo{address}{Cambridge, MA, USA}. p. \bibinfo{pages}{423–429}.
\bibitem[{Schneider et~al.(2009a)Schneider, Biehl and Hammer}]{petrabiehl}
\bibinfo{author}{Schneider, P.}, \bibinfo{author}{Biehl, M.},
  \bibinfo{author}{Hammer, B.}, \bibinfo{year}{2009}a.
\newblock \bibinfo{title}{Adaptive relevance matrices in learning vector
  quantization}.
\newblock \bibinfo{journal}{Neural Comput.} \bibinfo{volume}{21},
  \bibinfo{pages}{3532–3561}.
\newblock \URLprefix \url{https://doi.org/10.1162/neco.2009.11-08-908}.
\bibitem[{Schneider et~al.(2009b)Schneider, Biehl and Hammer}]{schneiderBiehl}
\bibinfo{author}{Schneider, P.}, \bibinfo{author}{Biehl, M.},
  \bibinfo{author}{Hammer, B.}, \bibinfo{year}{2009}b.
\newblock \bibinfo{title}{Distance learning in discriminative vector
  quantization}.
\newblock \bibinfo{journal}{Neural Computation} \bibinfo{volume}{21},
  \bibinfo{pages}{2942--2969}.
\newblock \URLprefix \url{https://doi.org/10.1162/neco.2009.10-08-892}.
\bibitem[{Schneider et~al.(2010)Schneider, Bunte, Stiekema, Hammer, Villmann
  and Biehl}]{schneider2010regularization}
\bibinfo{author}{Schneider, P.}, \bibinfo{author}{Bunte, K.},
  \bibinfo{author}{Stiekema, H.}, \bibinfo{author}{Hammer, B.},
  \bibinfo{author}{Villmann, T.}, \bibinfo{author}{Biehl, M.},
  \bibinfo{year}{2010}.
\newblock \bibinfo{title}{Regularization in matrix relevance learning}.
\newblock \bibinfo{journal}{IEEE Transactions on Neural Networks}
  \bibinfo{volume}{21}, \bibinfo{pages}{831--840}.
\bibitem[{Strobl et~al.(2007)Strobl, Boulesteix, Zeileis and
  Hothorn}]{strobl2007bias}
\bibinfo{author}{Strobl, C.}, \bibinfo{author}{Boulesteix, A.L.},
  \bibinfo{author}{Zeileis, A.}, \bibinfo{author}{Hothorn, T.},
  \bibinfo{year}{2007}.
\newblock \bibinfo{title}{Bias in random forest variable importance measures:
  Illustrations, sources and a solution}.
\newblock \bibinfo{journal}{BMC bioinformatics} \bibinfo{volume}{8},
  \bibinfo{pages}{25}.
\bibitem[{{Voggel} et~al.(2020){Voggel}, {Seth}, {Sand}, {Hughes}, {Strader},
  {Crnojevic} and {Caldwell}}]{voggel2020}
\bibinfo{author}{{Voggel}, K.T.}, \bibinfo{author}{{Seth}, A.C.},
  \bibinfo{author}{{Sand}, D.J.}, \bibinfo{author}{{Hughes}, A.},
  \bibinfo{author}{{Strader}, J.}, \bibinfo{author}{{Crnojevic}, D.},
  \bibinfo{author}{{Caldwell}, N.}, \bibinfo{year}{2020}.
\newblock \bibinfo{title}{{A Gaia-based Catalog of Candidate Stripped Nuclei
  and Luminous Globular Clusters in the Halo of Centaurus A}}.
\newblock \bibinfo{journal}{The Astrophysical Journal Supplement Series}
  \bibinfo{volume}{899}, \bibinfo{pages}{140}.
\newblock \DOIprefix\doi{10.3847/1538-4357/ab6f69},
  \href{http://arxiv.org/abs/2001.02243}{\tt arXiv:2001.02243}.
\bibitem[{Xiao et~al.(2020)Xiao, Cao, Fan, Costantin, Luo and
  Pei}]{xiao2020efficient}
\bibinfo{author}{Xiao, H.}, \bibinfo{author}{Cao, H.}, \bibinfo{author}{Fan,
  J.}, \bibinfo{author}{Costantin, D.}, \bibinfo{author}{Luo, G.},
  \bibinfo{author}{Pei, Z.}, \bibinfo{year}{2020}.
\newblock \bibinfo{title}{Efficient fermi source identification with machine
  learning methods}.
\newblock \bibinfo{journal}{Astronomy and Computing} , \bibinfo{pages}{100387}.

\end{thebibliography}
